\pdfoutput=1

\documentclass{elsart}
\usepackage{natbib}
\usepackage{amssymb}
\usepackage{graphicx,bbm,amsmath}

\newcommand{\U}[1]{\,{\rm{#1}}}
\newcommand{\euler}{\textrm{e}}

\newcommand{\I}[1]{_{\mathrm{#1}}}
\newcommand{\imag}{{\rm i}}
\newcommand{\mf}{\mathbf}

\newcommand*{\D}{\mathrm{d}}

\newcommand{\differential}{\>\textrm{d}}

\renewcommand{\vec}{\mf}

\newcommand{\au}{\U{a.u.}}

\newcommand{\Wcm}{\U{W/cm^2}}

\begin{document}

\begin{frontmatter}



\title{Frontiers of atomic high-harmonic generation}


\author{M. C. Kohler, T. Pfeifer,  K. Z. Hatsagortsyan, and C. H. Keitel}

\address{Max-Planck-Institut f\"ur Kernphysik, 
Saupfercheckweg 1, 69117 Heidelberg, Germany}

\tableofcontents

\begin{abstract}
In the past two decades high-harmonic generation (HHG) has become a key process in ultra-fast science due to the extremely short time-structure of the underlying electron dynamics being imprinted in the emitted harmonic light bursts. After discussing the fundamental physical picture of HHG including continuum--continuum transitions, we describe the experimental progress rendering HHG to the unique source of attosecond pulses. The development of bright photon sources with zeptosecond pulse duration and keV photon energy is underway. In this article we describe several approaches pointed toward this aim and beyond. As the main barriers for multi-keV HHG, phase-matching and relativistic drift are discussed. Routes to overcome these problems are pointed out as well as schemes to control the HHG process via alterations of the driving fields.  Finally, we report on how the investigation of fundamental physical processes benefits from the continuous development of HHG sources.

\end{abstract}


\end{frontmatter}


\section{Introduction}

 The photon can be seen in many regards as  an engine of technological progress  for the  21st century~\citep{NRC:LI-98} 
 --- a success story that began with the advent of lasers about 50 years ago~\citep{MAIMAN:LA-60}. Technology has continuously developed and lasers have become an important tool in industry and in almost every physics laboratory, independent of its field of research. Besides other parameters, the achievable laser intensities are a significant means to depict the progress in laser technology~\citep{MOUROU:OR-06}. The first lasers in the 1960s exhibited intensities well below $10^{10}\Wcm$. An enormous interest of the scientific community in this tool and vivid research led to rapid progress of the highest laser intensities  that can be  accomplished.
 Employing  chirped-pulse amplification~\citep{STRICKLAND:CP-85} renders it nowadays possible to create laser pulses
with peak intensities of $10^{22}\Wcm$~\citep{YANOVSKY:HI-08}. 
Intensities of more than $10^{24}\Wcm$ are envisaged for the upcoming {\it Extreme Light Infrastructure} (ELI) within the next decade, with probable impact on many areas in physics such as laser plasma acceleration, attoscience, warm dense matter, laboratory astrophysics, laser fusion ignition, nuclear and high-energy physics with lasers.
Besides the high peak intensities, pulse durations down to only a single cycle have also been achieved~\citep{NISOLI:CT-96,NISOLI:CT-97,SCHENKEL-CT03,WIRTH:ST-11}, the carrier envelope phase can be stabilized~\citep{BALTUSKA:CE-02,BALTUSKA:CE-03}, lasers have become more reliable and are available at a variety of wavelengths. 

With the invention of the laser, a bright coherent light source became available that had, among others,  huge impact on atomic physics opening the field of the laser spectroscopy and led to  a renaissance in the field of optics with the birth of non-linear optics~\citep{FRANKEN:SH-61,BOYD:NL-08}. 
A substantial advancement in non-linear optics 
and atomic physics
happened when the laser field strengths became comparable with the electric field experienced by a bound electron 
which was achieved in the 1980s. At that point, the non-perturbative regime of non-linear optics was entered. Fields with such a strength are able to significantly change the electronic dynamics of atoms and molecules. In particular,  large fractions of the electronic wave function can be transferred to the continuum within very short times. A number of new effects were discovered at that time.
First we mention above-threshold ionization (ATI)~\citep{AGOSTINI:AT-79} happening
when electrons ionized by an infrared (IR) laser field  were detected with a large kinetic energy well above the ionization threshold. 
Moreover, recollisions of those electrons with the core were observed, leading either to the knock-out of a second electron, the non-sequential double ionization~\citep{LHUILLIER:NS-83,FITTINGHOFF:NS-92}, or to  recombination along with the emission of extreme-ultra-violet (XUV) light, called high-order harmonic generation (HHG) -- the main process under consideration in this review.

\begin{figure}
\centering
\includegraphics[width=0.3\textwidth]{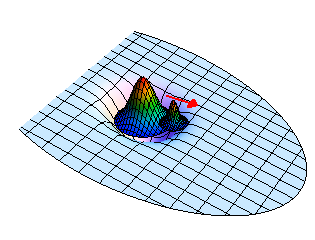}
\includegraphics[width=0.3\textwidth]{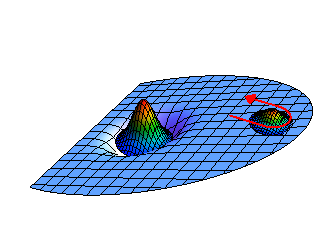}
\includegraphics[width=0.3\textwidth]{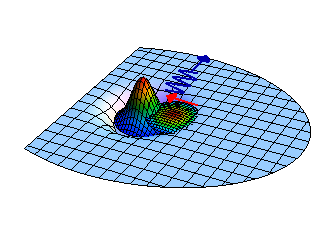}
\caption{\label{fig-3step}
(Color inline) Schematic of the three-step model for the HHG process. The blue surface is the superposition of the atomic binding potential and the electrostatic potential of the laser field. The bound and continuum part of the wave function are rainbow-colored. A tiny part of the wave function (continuum part) is just freed and subsequently driven away. When the laser field reverses its sign, the continuum part is stopped and accelerated back towards the core. At recollision both parts of the wave function interfere and give rise to coherent photon emission denoted by the blue wiggled line.}
\end{figure}

HHG was  discovered in 1987~\citep{McPherson:SM-87,FERRAY:MH-88}:  a rare gas illuminated by a laser emitted photons with an energy of several odd multiples of the laser frequency. Remarkably, the emission spectrum exhibits a plateau-like structure extending far until a sudden cutoff rather than a continuous exponential decrease typical for perturbative non-linear processes. This rapidly led to  intensive discussions about the physical origin of the phenomenon which eventually brought forth the three-step model (see Fig.~\ref{fig-3step}) 
discovered in 1993~\citep{Corkum:PP-93,SCHAFER:3S-93}. i) the electronic wave function of an atom is partially freed by a strong laser field, ii) then the ionized fraction is subsequently driven away in the continuum by the laser field, iii) finally, the wave packet is accelerated back to the ionic remnant, interfering with the bound part of the wave function giving rise to a strong, coherent, high-frequency response that can lead to the emission of a HHG photon along with the recombination of the electron into the bound state. 

HHG is fascinating both from being a fundamental example of nonperturbative laser-atom interaction dynamics 
and from a technological point of view, allowing creation of light flashes with exceptional properties: the energy of an emitted HHG photon is the sum of the binding energy and the kinetic energy acquired in the laser field. The photon energy can be extremely large reaching  several kilo-electronvolts (keV)~\citep{Seres:XA-06}.  In a world where technology is currently miniaturizing, x-rays with their tiny wavelengths may become the light of the future and thus HHG potentially of particular importance. Another property of HHG is its coherence, which arises because the phase of harmonic emission is locked to the laser phase~\citep{LEWENSTEIN:PH-95}, and under optimal conditions the harmonic yield scales quadratically with the number of contributing atoms. The last point indicates that the process cannot  be viewed only from a single-atom perspective. It is crucial for the emission that all atoms are phase-matched meaning that harmonics from different atoms have to emit with the same phase to allow for constructive interference between them. 
While the macroscopic aspect of HHG (phase-matching) is important to get a sizable harmonic yield and to create XUV/x-ray sources, the single-atom aspect is not less essential. Thus,
the harmonic spectrum contains structural signatures of the emitting atom or molecule which can be assessed by means of HHG spectroscopy~\citep{LEIN:AL-02,LEIN:RO-04,Itatani:TO-04,LEIN:NM-05,MORISHITA:MT-08}.

HHG opened the door for attoscience~\citep{Krausz:AP-09}. In fact, the bandwidth of emitted harmonics is large enough to allow 
for generation of light bursts much shorter than the laser pulse duration. 
The harmonic pulses~\citep{NISOLI:AS-09,CHANG:IS-10,Sansone:AS-11,CALEGARI2011}
can have durations down to the attosecond
regime~\citep{Sansone:IS-06,Goulielmakis:SC-08,KO:AS-2010}.
Attoseconds is the time scale of electron dynamics in atoms and ions. For a long
time, no tools have been available that were fast enough to resolve such
dynamics. For this reason, HHG has become a key process in atomic and molecular
physics and a state-of-the-art coherent XUV light source being nowadays
available in many research laboratories. With the emergence of extremely short
pulses from HHG, an entirely new pathway in exploring and controlling the
physics of atoms and
ions~\citep{SCRINZI:AP-06,KIENBERGER:AP-07,NISOLI:AS-09,Krausz:AP-09,BRIF:QC-10}
was opened.

The capability of the current HHG sources is limited to energies around a few
hundred eV and to pulse durations of several tens of attoseconds. An advancement
to shorter pulse durations and higher photon energies is highly desirable
because it will allow not only to track the electron dynamics in atoms but also
to time-resolve intra-nuclear dynamics. In this article we present selected
approaches to reach these aims. For previous reviews on similar topics, we refer
the reader
to~\cite{Protopapas:AP-97,Salieres:HH-99,Brabec:NO-00,Joachain:HI-00,
Becker:AT-02,AGOSTINI:PA-04,Pfeifer:RE-06,SALAMIN:RI-06,MOUROU:OR-06,
SCRINZI:AP-06,GAARDE:MA-08,Winterfeldt:OC-08,NISOLI:AS-09,Krausz:AP-09,
Popmintchev:AN-10,Sansone:AS-11,CALEGARI2011}.

An alternative technique for HHG is the irradiation of solids with super-strong lasers ($I>10^{16}\Wcm$) leading to the formation of overdense plasmas and the light emission from 
its surface~\citep{Lichters:OM-96,Quere:CW-06,DROMEY:DL-09,NOMURA:AP-09}. In our review we do not cover this approach and refer the reader to, e.g.,~\cite{TSAKIRIS:RI-06,Thaury:PL-10}.


\section{Fundamental concepts of HHG and attosecond pulses}

We begin with a brief presentation of the basic principles and the phenomenology of HHG along with an introduction to some particular theoretical concepts. This leads us to a discussion about the interference picture of HHG including continuum--continuum transitions. Further, we review milestone experiments that have led to the corresponding state-of-the-art experimental technology.

\subsection{Lewenstein model and Phenomenology}

The first theoretical description of HHG spectra has been accomplished by~\cite{Lewenstein:HH-94} on the basis of 
the strong-field approximation (SFA)~\citep{KELDYSH:IO-65,FAISAL:MA-73,REIS:EI-80}. 
In this work, a quasi-classical trajectory-based picture of the HHG process has
been established where the evolution of the wave function in the continuum,
derived from quantum mechanics, can finally be described by an ensemble of
classical trajectories. The theoretical model is developed as follows. The
harmonic emission is evaluated via the electric dipole moment (in atomic
units) \footnote{In atomic units the electron mass is $m\I{e}=1$, its charge
is $e=-1$, Planck's constant is $\hbar=1$ and consequently the speed of light
is found to be $c=137$.}
\begin{equation}
 \mf{d}(t)=-\langle\Psi(t)\vert \mf{x}\vert \Psi(t)\rangle \label{dipolemoment}\, ,
\end{equation}
where the electron wave function $\Psi(t)$ is determined within the SFA adopting  the single-active electron approximation~\citep{KULANDER:TM-88,Lewenstein:HH-94,PAULUS:PA-94}. The SFA is required for finding an analytic solution of the initially bound wave function evolving in the strong laser field. The fundamental SFA assumptions are that the binding potential is dominant before ionization and the laser field after ionization. This way, the Fourier transformed dipole moment $\tilde{\mf{d}}$ as a function of the harmonic frequency $\omega\I{H}$ was found to be
\begin{eqnarray}
\tilde{\mf{d}}(\omega\I{H})&=&\imag\int_{-\infty}^{\infty} \differential t\int_{-\infty}^t \differential t'\int \differential^3\mf{p} \langle \phi\I{0}\vert x\vert\mf{p}+\mf{A}(t)/c\rangle\nonumber\\
&&\;\;\;\;\;\;\;\;\times\langle\mf{p}+\mf{A}(t')/c\vert x E(t')\vert\phi\I{0}\rangle \euler^{-\imag(S(\mf{p},t,t')-I\I{p} (t'-t)-\omega\I{H} t)} \label{lewenstein-amplitude_spectr}\, ,
\end{eqnarray}
where
\begin{equation}\label{eq:action}
 S(\mf{p},t,t')=\tfrac{1}{2}\int_{t'}^t\,\differential \tau\bigl(\mf{p}+\mf{A}(\tau)/c\bigr)^2
\end{equation}
is the classical action of the laser field and $I\I{p}$ the binding potential, $\vert\phi\I{0}\rangle$ the bound state and $\vert\mf{p}\rangle$ the eigenstate of the momentum operator. $\mf{A}(t)$ and $\mf{E}(t)=-\partial_t\mf{A}(t)/c$ are the vector potential and electric field of the laser field, respectively.

The integrand in Eq.~\eqref{lewenstein-amplitude_spectr} is highly oscillating because of the complex argument of the exponent  and is numerically difficult to calculate exactly. Thus, one or several integrals can be carried out within the saddle-point approximation (see, e.g.,~\cite{Arfken:MM-05}) converting the quantum mechanical expression into a semi-classical expression. This means the integral is only evaluated around the stationary points of the phase $S(\mf{p},t,t')-I\I{p} (t'-t)-\omega\I{H} t$. These points are defined by the so-called saddle-point equations 
\begin{eqnarray}
\int^{t}_{t'}\differential t''[\mathbf{p}+\mathbf{A}(t'')/c]&=&0\label{0sp-cl}\\
~[\mathbf{p}+\mathbf{A}(t')/c]^2/2&=&-I\I{p} \label{0sp-ion}\\
~[\mathbf{p}+\mathbf{A}(t)/c]^2/2+I\I{p}&=&\omega\I{H}\label{0sp-rec}
\end{eqnarray} 
which correspond to energy conservation at ionization [Eq.~\eqref{0sp-ion}] and recollision [Eq.~\eqref{0sp-rec}] and to the recollision condition of the classical electron [Eq.~\eqref{0sp-cl}]~\citep{Lewenstein:HH-94}.
The stationary points or saddle points themselves represent the ionization and recollision time of the considered classical trajectory as well as the canonical momentum. It turns out that it is sufficient to sum over a small number of classically allowed trajectories for each energy to calculate the spectrum given by Eq.~\eqref{lewenstein-amplitude_spectr}.  For more details about the semi-classical picture see, e.g.,~\cite{Salieres:HH-99,MILOSEVIC:SM-01,Becker:AT-02} and for its relativistic extension~\cite{MILOSEVIC:UH-00,MILOSEVIC:UH-02,SALAMIN:RI-06,KLAIBER:LO-07}. Apart from the analytic approach above, the laser-atom-dynamics governed by the Schr\"odinger equation or relativistic wave equation can also be solved numerically~\citep{MULLER:NS-99,BAUER:QP-06,MOCKEN:DE-08,RUF:KG-09}. 
Moreover, an approach based on the SFA but beyond the semi-classical picture has been developed recently~\citep{PLAJA:CC-07}.

\begin{figure}[h]
\centering
\includegraphics[width=0.45\textwidth]{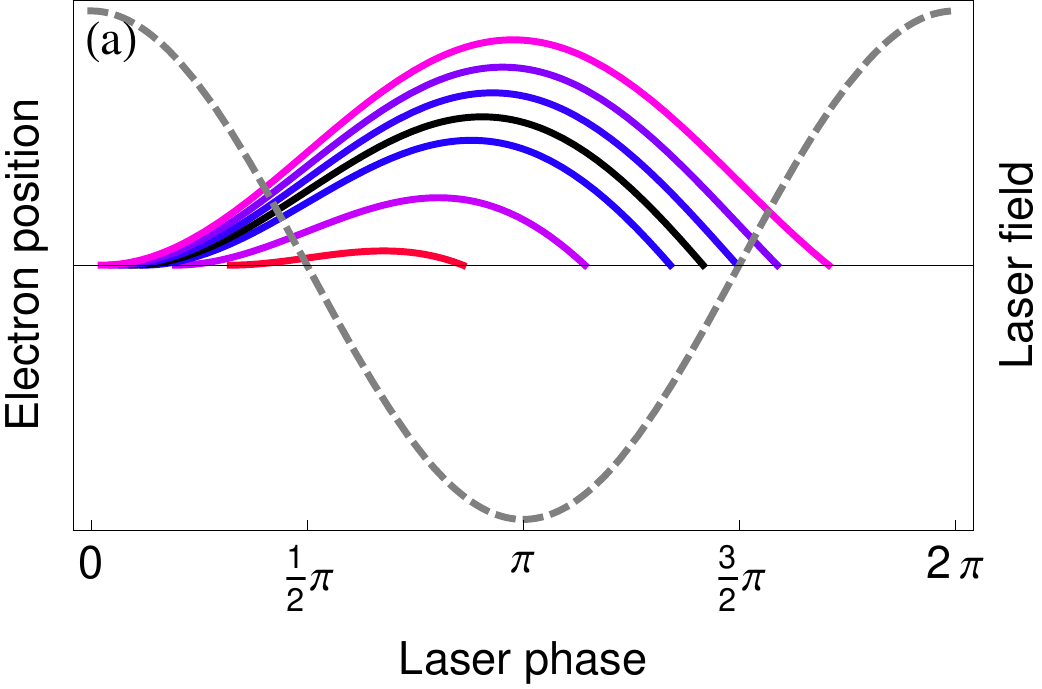}\hskip 0.1cm
\includegraphics[width=0.45\textwidth]{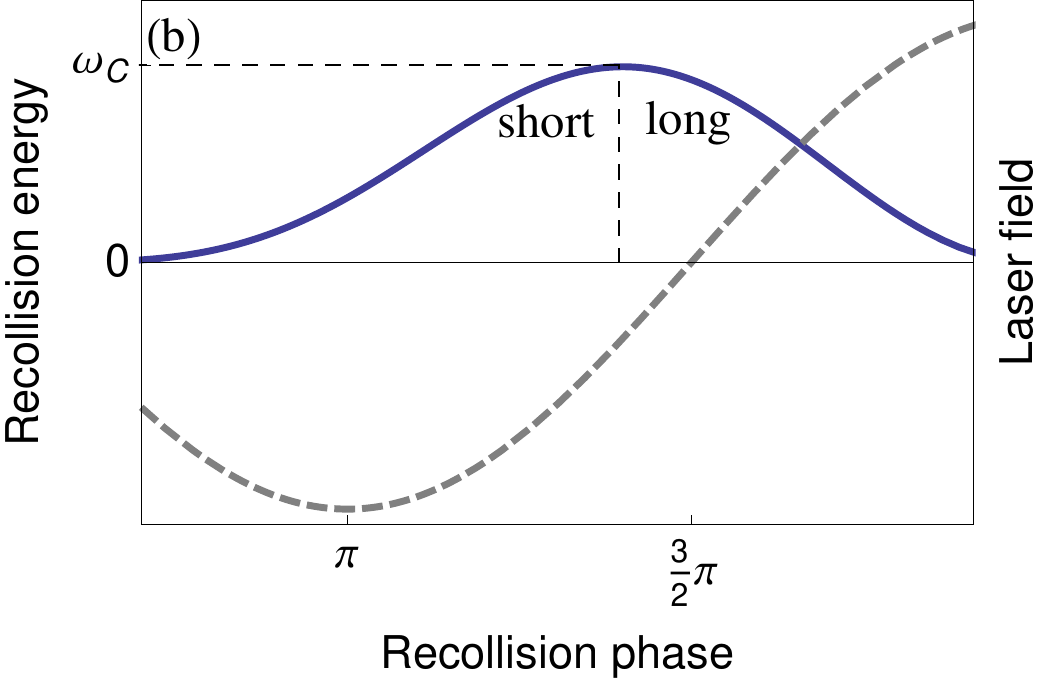}
\caption{\label{fig-classical-traj}
(Color online) The 
gray dashed line shows the laser field in both parts of figure. (a) The solid
colored lines are different classical trajectories.  The trajectories start at
different phases in the laser field. The recollision energy is encoded in their
color. The energy increases from red to blue and the trajectory with the maximum
energy (cutoff) is marked in black.  In (b) we see the energies of the
recolliding trajectories for different recollision times as solid line.}
\end{figure}

The semi-classical picture has opened the perspective to intuitively understand a wide range of problems connected with HHG.
Inspecting the classical trajectories, we briefly explain several principles of HHG. In Fig.~\ref{fig-classical-traj}~(a), we show different classical trajectories (colored lines) together with the laser field (gray dashed line). Recollision is only possible for trajectories ionized in the quarter cycle after each maximum or minimum of the laser cycle. Trajectories starting before do no re-encounter the origin. All trajectories originating from the same half cycle recollide at different times and  have different recollision energies. This also means that different harmonic wavelengths are emitted at different times and, thus, the emitted light has an intrinsic chirp, the so-called attochirp~\citep{MAIRESSE:AS-2003}. The recollision energies versus the recollision phase are shown in Fig.~\ref{fig-classical-traj}~(b). It can be seen that each energy is emitted twice per half cycle. These two branches of trajectories are termed {\it long} and {\it short} according to their excursion time in the continuum. 
As indicated in Fig.~\ref{fig-classical-traj}~(b), there is one trajectory with a maximum energy $\omega\I{c}$. This energy is called the cutoff energy and can be derived from the classical equations of motion yielding~\citep{Corkum:PP-93} 
\begin{equation}
 \omega\I{c}=3.17\,U\I{p}+I\I{p}\, , \label{eq:cutoff}
\end{equation}
where $I\I{p}$ is the ionization energy and $U\I{p}=\frac{E\I{0}^2}{4\omega^2}$ is the ponderomotive potential equal to the average quivering energy of an electron in a sinusoidal laser field with peak strength $E\I{0}$ and frequency $\omega$.

\begin{figure}
\centering
\includegraphics[width=0.45\textwidth]{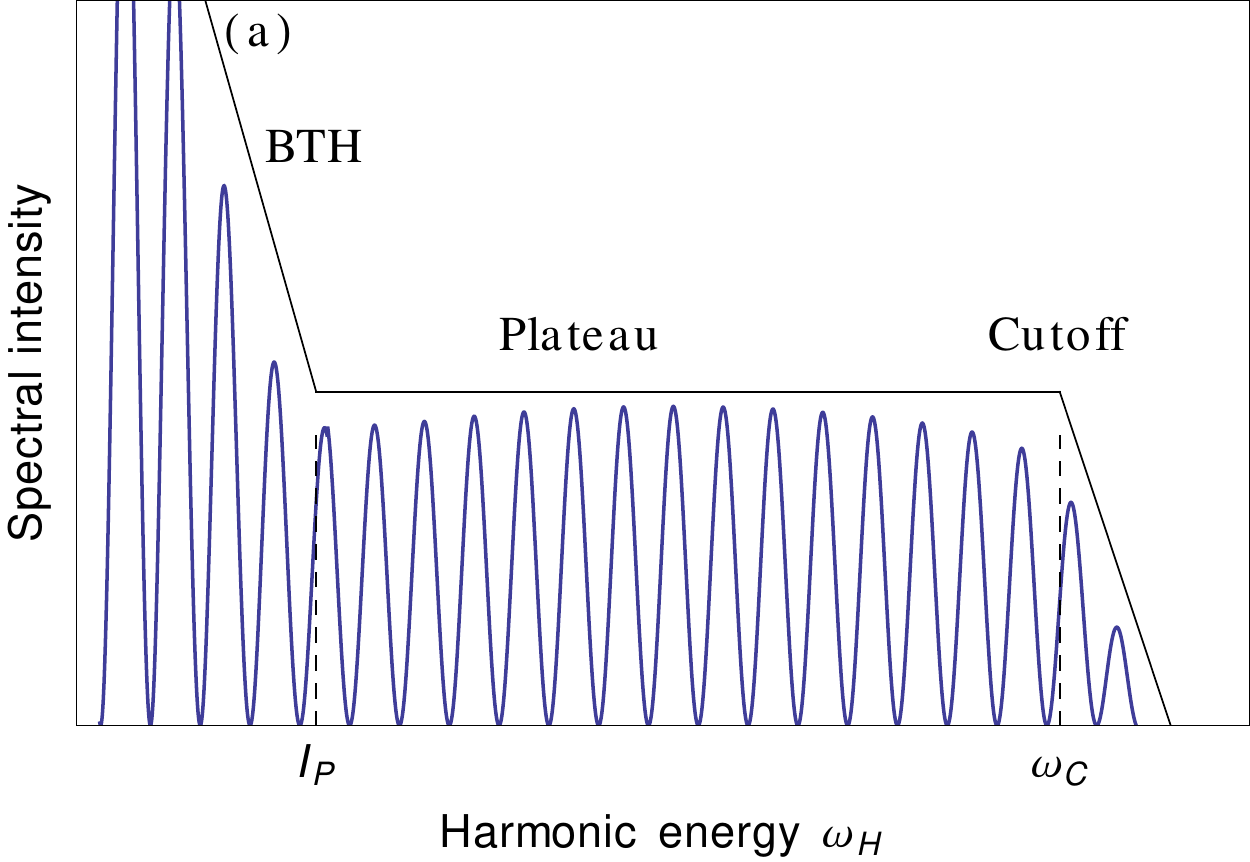}\hskip 0.3cm
\includegraphics[width=0.45\textwidth]{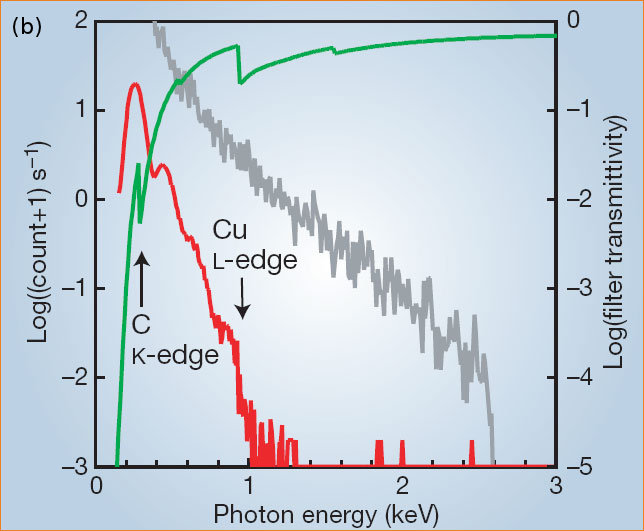}
\caption{\label{fig-hhg-typical}
(Color online) (a) Schematic of a typical HHG spectrum. The below-threshold harmonics (BTH)
exhibit an exponential decrease in the first part followed by the plateau and a
steep drop after the cutoff. (b) Experimentally measured HHG spectrum (red
line, lowest line for high energies), transmittance function of the various
filters (green, highest line for high energies) and the theoretical
single atom emission yield (gray, diagonal line).  Figure reprinted with
permission from~\cite{SERES:SC-05}.
Copyright 2005 by Nature (London).}
\end{figure}

 A typical HHG spectrum is shown in Fig.~\ref{fig-hhg-typical}~(a).  The part of the spectrum below $I\I{p}$ is called below-threshold harmonics (BTH).  In the early years of laser physics~\citep{FRANKEN:SH-61} this branch of the spectrum has been modelled by taking the laser field perturbativly into account but it is under active discussion at the moment~\citep{HU:NT-01,YOST:BT-09,POWER:NT-10,Hostetter:BT-10,Soifer:BT-10,Liu:BT-11}.  The  long plateau in the spectrum ending at the cutoff energy $\omega\I{c}$ originates from a superposition of all classically allowed trajectories. The red line in (b) shows a state-of-the-art experimentally measured HHG spectrum ranging to keV photon energy, from~\cite{SERES:SC-05}. Macroscopic effects are responsible for the deviation between (a) and (b). Moreover, the single harmonics cannot be resolved for such a large bandwidth.

Unfortunately, HHG has a poor conversion efficiency. This is mainly because the free wave packet undergoes quantum spreading during the continuum motion. When the wave packet is driven back, most parts of the wave function miss the core due to its large spatial extent after it has spread out. The impact of the spreading can be easily estimated. The dimensions of the bound wavefunction are typically on the order of $1\au$. The transversal spreading velocity can be estimated by $v\I{\perp}=\sqrt{3E}/(2I\I{p})^{1/4}$~\citep{Popov:TI-04,IVANOV:CO-96}. For a typical HHG experiment ($E\sim 0.1\au$, $I\I{p} \sim1\au$ and $\omega\I{L}=0.057\au$, where $\omega\I{L}$ is the laser frequency), we find a wave packet dimension of $x=v\I{\perp}\frac{2\pi}{\omega\I{L}}=30\au$ at recollision. Thus, spreading in the two perpendicular dimensions with respect to the laser polarization axis reduces the HHG yield by a factor of $(30/1)^2\sim10^3$.

Let us come back to the time structure of HHG. Around each maximum of the laser field, portions of the wave function tunnel out, recollide after a fraction of the laser period [see classical trajectories in Fig.~\ref{fig-classical-traj}~(a)] and lead to the formation of a light burst. In a periodic laser field the process repeats itself each half cycle of the driving field and the emitted field $\tilde{E}\I{H}(\omega\I{H})$  only differs in a phase-shift of $\pi$ and the sign. Thereby it is assumed that two adjacent half cycles of the laser field are always identical apart from their sign and the atom possesses a central symmetry and thus the continuum dynamics are inverted after each half cycle leading to the difference in the sign of the dipole moment.  A typical spectrometer being slow on the timescale of the laser period would thus measure a superposition of all these emitted pulses~\citep{PROTOPAPAS:HG-96}
\begin{equation}
\tilde{E}\I{H}(\omega\I{H})\euler^{\imag\frac{\omega\I{H}}{\omega} \pi}-\tilde{E}\I{H}(\omega\I{H})\euler^{\imag\frac{\omega\I{H}}{\omega}2 \pi}+\tilde{E}\I{H}(\omega\I{H})\euler^{\imag\frac{\omega\I{H}}{\omega} 3\pi}+\ldots=\tilde{E}\I{H}(\omega\I{H})\sum_j (-1)^j \euler^{\imag\frac{\omega\I{H}}{\omega}j \pi}
\end{equation}
with significant contributions to the spectrum only at odd multiples of the laser frequency as in Fig.~\ref{fig-hhg-typical} even though the individual pulses exhibit a continuous spectrum. That is the physical reason for calling the effect high-harmonic generation. When violatating the former assumptions, even-order harmonics can also also be generated,~e.g.,~in two-color laser fields~\citep{Kondo:EH-96}, via asymetric molecules~\citep{Gavrilenko:EH-00,Kreibich:EH-01} or driving pulses short enough that the periodicity is broken~\citep{Spielmann:GC-97}.
Experimental progress  meanwhile allows for the restriction of the HHG process to a half cycle only (see Sec.~\ref{EXP-SOURCE} for more details). In this case a continuous spectrum is measured. The duration of such a pulse ($\Delta t$) depends on two issues: first, the available spectral bandwidth ($\Delta \omega\I{H}$) of the harmonics dictates the shortest possible pulse duration via the Fourier limit. It can be expressed by the time-bandwidth product
\begin{equation}\label{eq:time-bandw}
 \Delta \omega\, \I{H}\Delta t =3.6\U{as}\,\U{keV}
\end{equation}
under the assumption of  a plateau-like spectrum. Second, the Fourier limit or bandwidth limit can only be reached when the pulse has no chirp. Thus, the intrinsic attochirp has to be compensated to compress the harmonic pulse toward its fundamental limit.

\subsection{Interference model of HHG}\label{sec:intf-model}

One of the most fundamental properties of harmonic light flashes  is their coherence, i.e., the phase of the harmonic light emitted from a single atom is locked to the laser phase and under suitable conditions even the phases of the harmonics from different atoms match. This particular property makes HHG a unique process with a variety of applications, for instance, rendering  attosecond pulse generation possible. In order to understand why the light can be coherently emitted, the generation process of a single atom has to be seen from a quantum mechanical perspective: For a typical HHG scenario in the tunnel ionization or multiphoton regime~\citep{Protopapas:AP-97}, the initially bound wave function is partially ionized. The fraction of the wave function promoted to the continuum is then driven in the strong laser field and can eventually recollide. At that point, both, the ionized and the bound parts of the wave function interfere within the binding potential giving rise to a strong, coherent high-frequency dipole response that can lead to the emission of an HHG photon along with the recombination of the electron into the bound state~\citep{Pukhov:TS-03,Itatani:TO-04}. We term the process continuum--bound (CB) HHG because a continuum wave packet and a bound wave packet are involved.

The interference picture of HHG can also be illustrated more quantitatively. In an electron--atom collision, the coherent part of the emitted radiation~\citep{LAPPAS:SL-93,BURNETT:CB-92} can be calculated via the expectation value of the acceleration using the Ehrenfest theorem
\begin{equation}
 {\mathbf a}(t)=-\langle \Psi(t)\vert \nabla V\vert \Psi(t)\rangle
\label{a_general}
\end{equation}
with the ionic potential $V$. Suppose the wave function would be a superposition of a recolliding plane wave with momentum $\mf{p}$ and the bound state $\vert \phi\I{0} \rangle$ as in the usual HHG scenario: $\vert \Psi(t)\rangle=a\I{1}\vert \mf{p} \rangle \euler^{-\imag \frac{\mf{p}^2}{2} t}+a\I{2}\vert \phi\I{0} \rangle \euler^{\imag I\I{p} t}$. The resulting acceleration  $a(t)=-a\I{1}^* a\I{2} \langle\epsilon\I{1}\vert \nabla V\vert\epsilon\I{2}\rangle \euler^{-\imag (\frac{\mf{p}^2}{2}+I\I{p}) t}+\textrm{c.c.}$ would both oscillate and lead to emission at the expected frequency $\omega\I{H}=\frac{\mf{p}^2}{2}+I\I{p}$. In the case, the bound wave function has been depleted ($a\I{2}=0$), the oscillations along with the coherent HHG emission are absent~\citep{Pukhov:TS-03}.  This situation occurs for  laser fields strong enough to ionize the whole wave function before recollision. The intensity regime is called saturation or over-the-barrier ionization (OBI) regime and is reached when the barrier-suppression field strength
$ E\I{BSI}=\frac{I\I{p}^2}{4 Z}$~\citep{AUGST:OB-89}
is exceeded, where $Z$ is the residual charge seen by the ionized electron. 

The former simple model based on eigenstates holds strictly true only for time-independent Hamiltonian but its picture remains true in atomic HHG with collision times typically short on the time scale of a laser period. Note that apart from coherent radiation considered in this review, a recolliding electronic wave packet can also cause emission of incoherent radiation due to spontaneous recombination. 
This spontaneous process also occurs when the electronic wave function has been fully depleted.
The phase of the spontaneous recombination  amplitude is random because of the arbitrary phase of the final state (ground state).
However, in a macroscopic gas target this kind of emission yield scales only linearly with the number of atoms instead of the quadratic scaling typical for the coherent CB transition in the phase-matched case, in addition to it being non-directional as compared to the directed CB HHG light.

Needless to say that further possibilities exist to create a time-dependent acceleration Eq.~\eqref{a_general} leading to coherent photon emission.
Especially relevant in HHG is the presence of two continuum wave-packet contributions within the range of the ionic potential. In this case, the wave function can be illustrated as
\begin{equation}\label{eq:cc-wf}
 \vert\Psi\I{1}(t)\rangle=a\I{1} \vert p\I{1}\rangle \euler^{-\imag p\I{1}^2/2 t}+a\I{2} \vert p\I{2}\rangle \euler^{-\imag p\I{2}^2/2 t}
\end{equation}
leading to the acceleration
$a(t)=-a\I{1}^* a\I{2} \langle p\I{1}\vert \nabla V\vert p\I{2}\rangle \euler^{-\imag (p\I{2}^2-p\I{1}^2)/2 t}+\textrm{c.c.}$ that is beating and coherently emitting 
Bremsstrahlung at the difference energy $(p\I{2}^2-p\I{1}^2)/2$ between both wave-packet fractions. 
This class of transitions is termed continuum--continuum (CC) HHG and it has already been  noticed in the early years of HHG~\citep{Lewenstein:HH-94}. 
Although it was confirmed by several authors in the past~\citep{BECKER:UN-97,KUCHIEV:HG-99,MILOSEVIC:SM-01,MILOSEVIC:RC-03}  that CC HHG has a magnitude of several orders of magnitude lower than CB HHG, interest in this transition has increased recently~\citep{PLAJA:CC-07,HERNANDEZ:CC-07,Hernandez:sw-09,KOHLER:CC-10} 
as ways have been found to separate it from the generally stronger CB HHG.

On the bottom line of the former discussion, a unified picture for HHG including the CB and all CC transitions can be formulated: the interference of two parts of the same wave function with different energies in a potential causes coherent photon emission at their difference energy either due to recombination radiation or due to Bremsstrahlung~\citep{KOHLER:CC-10}.
For instance, in the case of CB HHG, a recolliding electronic plane wave (energy $\epsilon\I{1}$) interferes with the bound wave packet (energy $-I\I{p}$) leading to photon emission at the difference energy ($\epsilon\I{1}+I\I{p}$).

\subsection{Continuum--continuum HHG}

\begin{figure}
 \centering
\includegraphics[width=4.0cm]{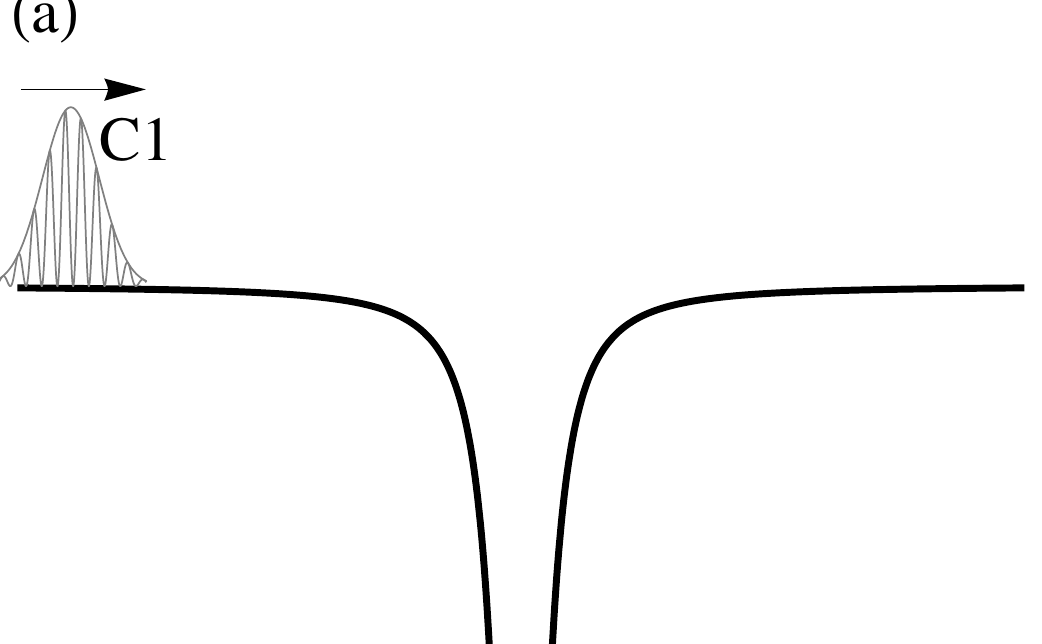} 
        \hspace{0.4cm}
\includegraphics[width=4.0cm]{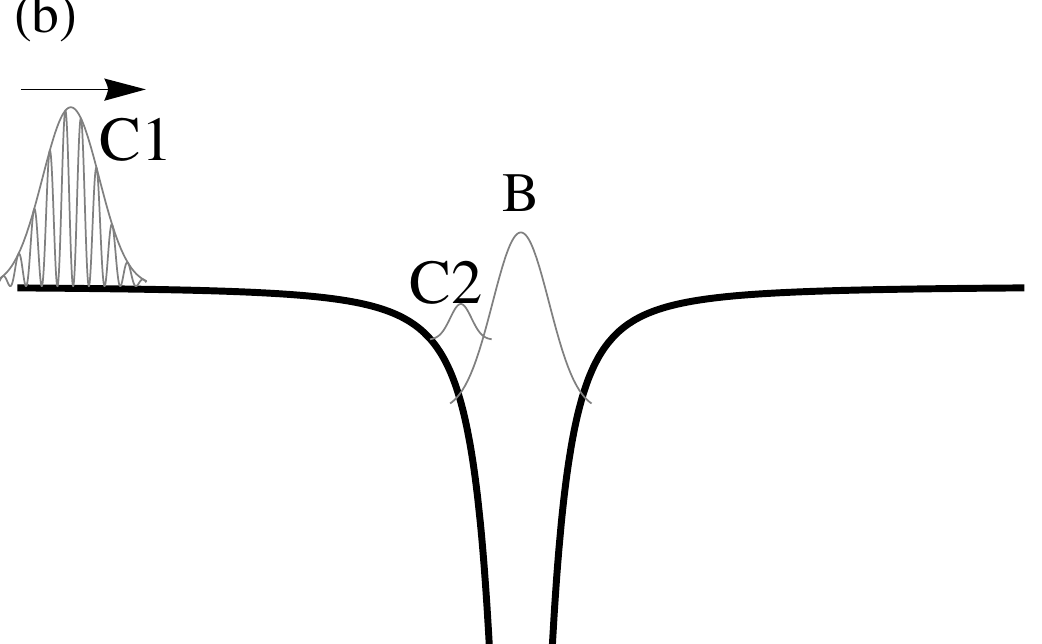}
        \hspace{0.4cm}
\includegraphics[width=4.0cm]{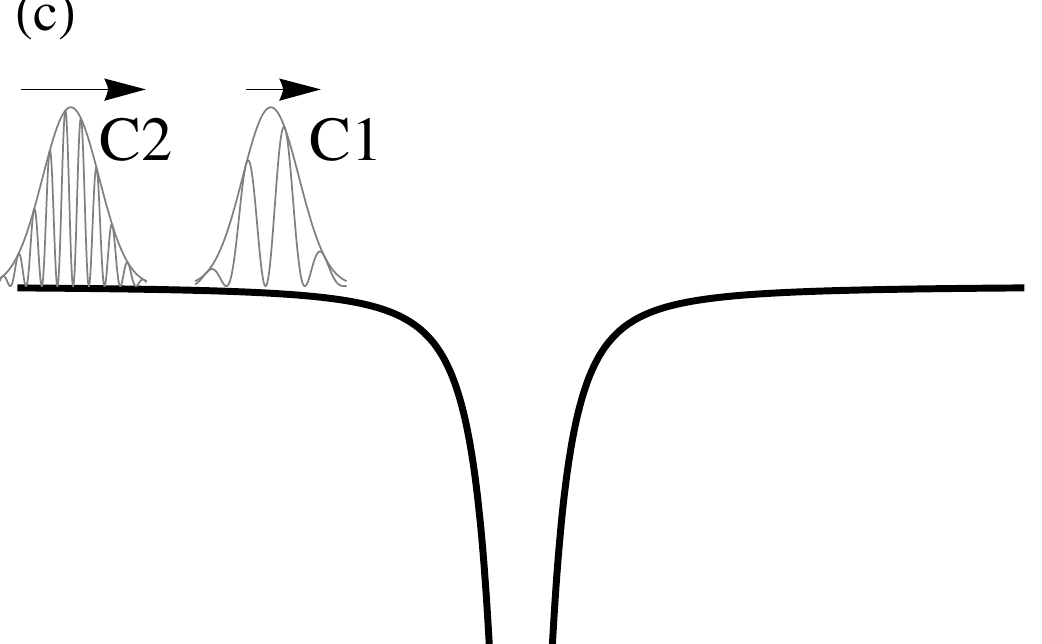}

\caption{
 Schematic of the different CC transitions occurring in HHG. We
concentrate on the coherent response. The thick black line is the potential, the thin 
gray lines are the different wave-packet portions. In (a)  Bremsstrahlung
is emitted when a single wave packet (C1) interacts with the core. In (b) the
recolliding wave packet will interfere with the bound wave packet (B) and a
just-emerging continuum wave packet (C2) leading to  CB and
CC harmonics, respectively. In (c), two continuum wave packets (C2 and C1) of {\it different momentum} recollide at the
same time and interfere with each other emitting photons of exactly their
kinetic-energy difference.}\label{fig:ccHHGs}
\end{figure}

As mentioned in the previous section, CC HHG has a several orders lower magnitude than CB HHG and, hence, its experimental discrimination is demanding and has not been accomplished yet. The reason for this is the smallness of the free-free transition cross-section ($\sim r\I{e}^2$, with the electron classical radius $ r\I{e} $) with respect to the free-bound one ($\sim r\I{B}^2$, with the Bohr radius $r\I{B}$). In the following we distinguish the three different CC transitions described in the literature and discuss new results which show the pathway to separate CC HHG from CB HHG.  All CC transitions share that interference between two continuum wave packets  within the range of the binding potential [see Fig.~\ref{fig:ccHHGs}] is responsible for the emission of coherent Bremsstrahlung. The scenarios can be distinguished by the preceding dynamics of the involved wave packets.

\subsubsection{CC HHG with a single continuum wave packet}\label{sec:CCbrems}

The schematic in Fig.~\ref{fig:ccHHGs}~(a) displays the recollision of a single wave packet with the ionic core. Since the ground state is not populated, no coherent (CB) emission takes place at the sum of the kinetic energy and the binding energy as argued before. However, due to the nonzero spectral width of the wave packet, the requirement of at least two simultaneously recolliding different momentum components is met [see Eq.~\eqref{eq:cc-wf}]. The recolliding electronic wave packet scatters at the ionic core which results in the 
transition between the two momentum states available in the electron wave packet along with photon emission forming coherent Bremsstrahlung~\citep{PROTOPAPAS:HG-96,Watson:WR-97,MILOSEVIC:RC-03}.
The availability of ground-state population is  irrelevant for this process and, therefore, it also occurs  for laser intensities in the saturation regime where the whole electronic wave function is completely ionized.

Very short radiation bursts down to the attosecond regime can even be emitted by this mechanism as calculations by \citet{EMELIN:HG-05,EMELIN:SB-08} show.
For this purpose, strong laser fields are required to ionize the electron within an extremely short time interval. 
In this case, the electron wave function is a single compact continuum wave packet after ionization. Later at recollision, the overlap between the wave function and the ion along with the emission process via the CC-transitions is confined to an attosecond time window under the precaution that the wave packet spread is within appropriate limits.  According to the finite spectral width of the recolliding wave packet, the spectral range of the discussed CC transitions is limited and the photon energies are low compared to the kinetic energy at recollision and CB HHG.

\subsubsection{CC HHG with a Rydberg state}\label{sec:CCio}
Another kind of CC transition [shown in Fig.~\ref{fig:ccHHGs}~(b)] happens when a recolliding wave packet interferes with a just-ionizing part of the bound wave function~\citep{Lewenstein:HH-94,BECKER:UN-97,KUCHIEV:HG-99,MILOSEVIC:SM-01,MILOSEVIC:RC-03, PLAJA:CC-07,HERNANDEZ:CC-07,Hernandez:sw-09} within the range  of the ionic potential. As CB HHG, it only occurs in the saturation regime (with $E<E\I{BSI}$) where the atom is not depleted. 
This CC transition always accompanies CB HHG and is 
even
indistinguishable from the CB transition into an excited Rydberg state where the latter has been populated from the initial ground state in the laser field. 
Only the energy of the CC transition is slightly smaller than that of the CB transition. Thus, the emission spectrum has similarities imprinted: the CC spectrum exhibits the same plateau-like structure as the CB spectrum, see~\cite{MILOSEVIC:SM-01}. 

Due to the weakness of the CC spectrum with respect to CB HHG, it can be neglected in many cases~\citep{KUCHIEV:HG-99,MILOSEVIC:SM-01}. However, recent results~\citep{PLAJA:CC-07} indicate significant CC contributions in the multiphoton regime. In this work, the ratio between the CC and CB amplitude was estimated to be $1-1/(4\gamma^2)$ with the Keldysh parameter $\gamma=\sqrt{I\I{p}/(2U\I{p})}$.

\subsubsection{CC HHG with two continuum wave packets}
\begin{figure}[h]
\centering
\includegraphics[width=9.0cm]{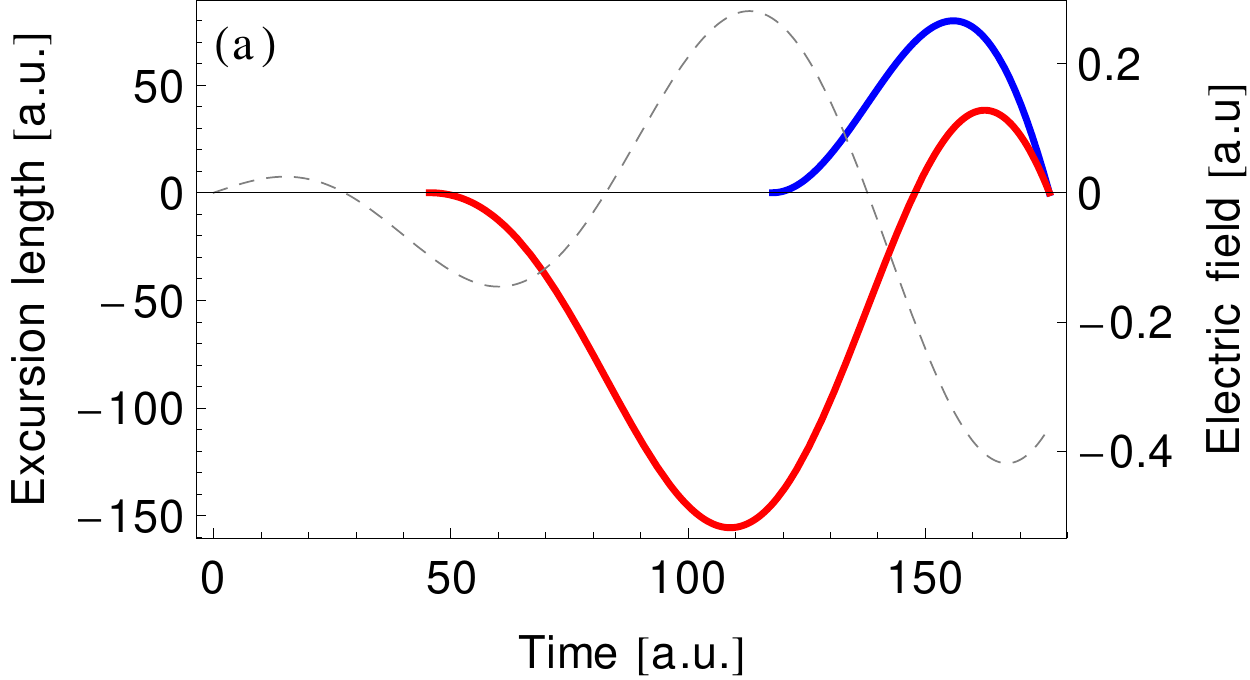}
\caption{\label{fig:cc-traj}
(Color online) Classical electron trajectories (solid lines) evolving in a multi-cycle laser field (dashed line).}
\end{figure}

In what follows, we consider two distinct continuum wave packets recolliding at the same time as sketched in Fig.~\ref{fig:ccHHGs}~(c). In contrast to the transitions in Section~\ref{sec:CCbrems} and~\ref{sec:CCio},  both wave packets have been ionized in two different half cycles and have been evolved in the continuum for some time. Thus, both wave packets recollide with a kinetic energy that is nonzero and comparable with the ponderomotive potential of the laser field. The scenario can occur in multi-cycle laser pulses as sketched in Fig.~\ref{fig:cc-traj} where classical electron trajectories are shown (solid lines). Two portions of the wave function ionized at the starting times of the two trajectories will recollide simultaneously and result in CC emission at their kinetic  energy difference.

\begin{figure}[h]
\centering
\includegraphics[width=12.0cm]{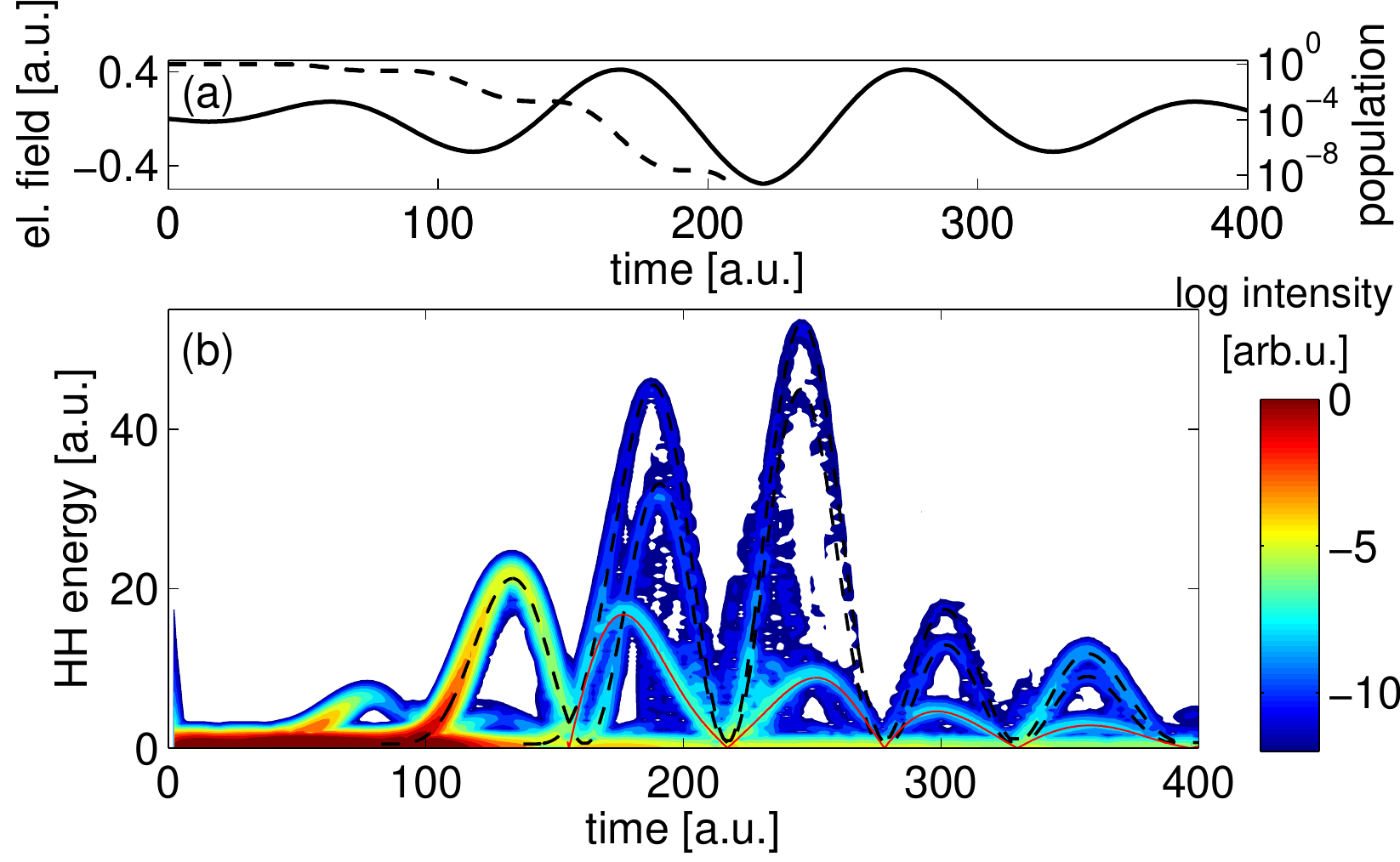}
\caption{\label{fig-HHGnumerical}
(Color online) Time-frequency analysis of HHG showing the signature of CC wave-packet interference.  a) The laser pulse used for the calculation (solid line, left axis) and ground-state population (dashed line, right axis).  b)~The windowed Fourier transform of the acceleration Eq.~\eqref{a_general}.  The two dashed black lines represent the classically calculated kinetic energies of electrons returning to the ion and the solid red line represents their difference energy. Figure reprinted with permission from~\cite{KOHLER:CC-10}. Copyright 2010 by the American Physical Society.}
\end{figure}

In contrast to the CC transition in Section~\ref{sec:CCio}, this transition also occurs when the wave function is fully depleted at the recollision moment
and, thus, can play a dominant role for HHG in the saturation regime.
This regime has been studied by means of hydrogen atoms in~\cite{KOHLER:CC-10} by a numerical solution of the three-dimensional time-dependent Schr\"odinger equation 
using the \textsc{Qprop} of~\citet{BAUER:QP-06}. The laser pulse is shown in Fig.~\ref{fig-HHGnumerical}~(a) and is chosen such that almost complete depletion of the ground state occurs on the leading edge of the pulse [see dashed line in Fig.~\ref{fig-HHGnumerical}~(a)]. 
To analyze the time-resolved  frequency response of HHG, the windowed Fourier transform of the acceleration expectation value obtained from the TDSE calculation is calculated and displayed in Fig.~\ref{fig-HHGnumerical}b. For comparison, the two dashed black lines in the figure display the classical recollision energies for trajectories starting from two different laser half cycles [first two peaks of the laser pulse], respectively, which are in agreement with the traditional CB signal whereas the red line represents their difference.
The CC transition is evidenced by the excellent agreement of the quantum-mechanical response with the red line. Interestingly, the CC component of the dipole response is the dominant contribution for several half cycles after $t=150\au$. This can be understood from the fact that depletion of the ground state occurs around that time. Then, coherent HHG can only occur by the presence of the various parts of the wave function in the continuum.

\begin{figure}
 \centering
\includegraphics[width=9cm]{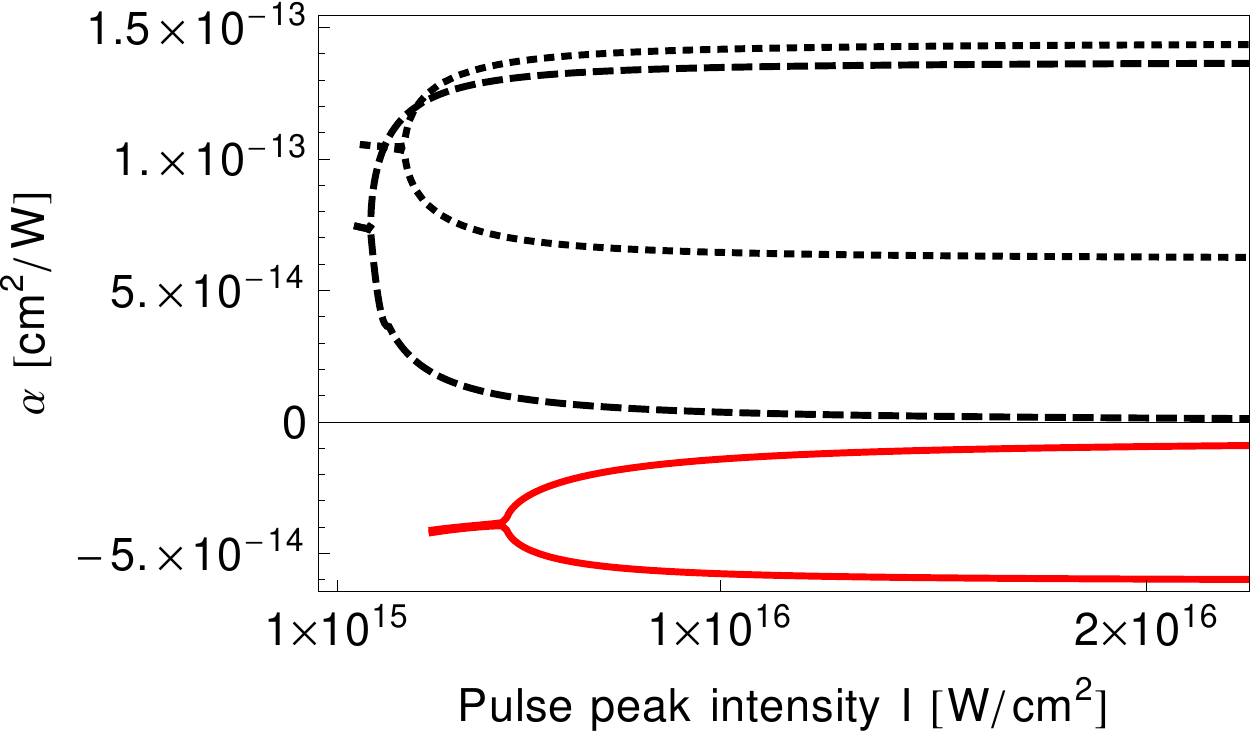}
\caption{(Color online) Intensity dependence of $\alpha_{\rm CB}$ and $\alpha_{\rm CC}$  for the six quantum orbits at the harmonic frequency $\omega\I{H}=8$ a.u. with respect to the peak pulse intensity which was chosen to be $10^{16}\Wcm$ in the example of Fig.~\ref{fig-HHGnumerical}. The solid red lines represent the two CC contributions. The dashed black line depicts the values of $\alpha_{\rm CB}$ for the CB long and short contributions with a single return. The dotted black line denotes the CB contributions that arise from trajectories emitting at the second return.}\label{fig:cc_phase-match}
\end{figure}

Moreover, the CC transition can be described within a strong-field approximation  model suitable for the OBI regime~\citep{KOHLER:CC-10}. The model is based on the evaluation of the  acceleration  $\mf{a}(t)=-\langle \Psi (t)\vert \mf{\nabla} V\vert \Psi (t)\rangle$~\citep{GORDON:RC-05} rather than the dipole moment to include the distortion of the recolliding waves by the Coulomb potential required for momentum conservation. The saddle-point approximation is applied to the expression rendering a computationally fast evaluation of the process possible.
This analytical model also allows to extract the HHG dipole emission phase
\begin{equation}
\phi(I)=\alpha\I{i}I\,, \,\,\, (i={\rm CB,CC})
\label{phase}
\end{equation}
 being a function of the laser pulse peak intensity $I$~\citep{LEWENSTEIN:PH-95}.
The results are shown in Fig.~\ref{fig:cc_phase-match} for a photon energy of
$\omega\I{H}=8$~a.u. emitted between $t=160\au$ and $t=210\au$ where 6 different
contributions exist [see Fig.~\ref{fig-HHGnumerical}b]. The dashed and dotted
lines represent  $\alpha_{\rm CB}$ for the usual CB transitions with
single and multiple return.
The red solid line displays the intensity dependence of $\alpha_{\rm CC}$ for the CC transition. 
The result of Fig.~\ref{fig:cc_phase-match} reveals a striking difference for CB as compared to CC transitions: the sign of $\alpha\I{CC}$ and $\alpha\I{CB}$ differs for both types of transitions, consequently, the phase-matching conditions for macroscopic propagation will be different for these transitions. 
The different phase-matching behavior could allow for the discrimination of the CC harmonics from the CB harmonics after propagation through the medium. 

The intensity of CC HHG depends on the effective atomic or molecular (ionic) potential mediating the transition between the two continuum electron wave packets. This feature can be
employed for qualitatively advancing tomographic molecular imaging~\citep{Itatani:TO-04}: instead of probing the orbital shape of the active electron, the effective atomic or molecular potential could be assessed by investigation of CC spectra. 

\subsection{Experimental advances} \label{EXP-SOURCE}
Since the first discovery and experiments on high-order harmonic generation
(HHG) in the late 1980s~\citep{McPherson:SM-87,FERRAY:MH-88}, a continuous and
still ongoing revolution in experimental technology has lead to progressively enhanced HHG and attosecond light sources as well as to the remarkable understanding and
control over this nonlinear-optical light conversion process.  The experimental
setup is conceptually simple and involves only the focusing of a sufficiently
high-energy ultrashort laser pulse (driver) into a gaseous medium.  HHG takes
places in each atom microscopically as discussed in the previous sections, and
macroscopic phase matching of the individual emitters throughout the interaction
region (which typically has lengths ranging from sub-mm to few cm and diameters
in the few to hundreds of micron regime) ensures the generation of a bright
coherent beam of harmonics after the conversion medium, typically copropagating
with the generating fundamental driver pulse.

\begin{figure}
\centering
\includegraphics[width=8cm]{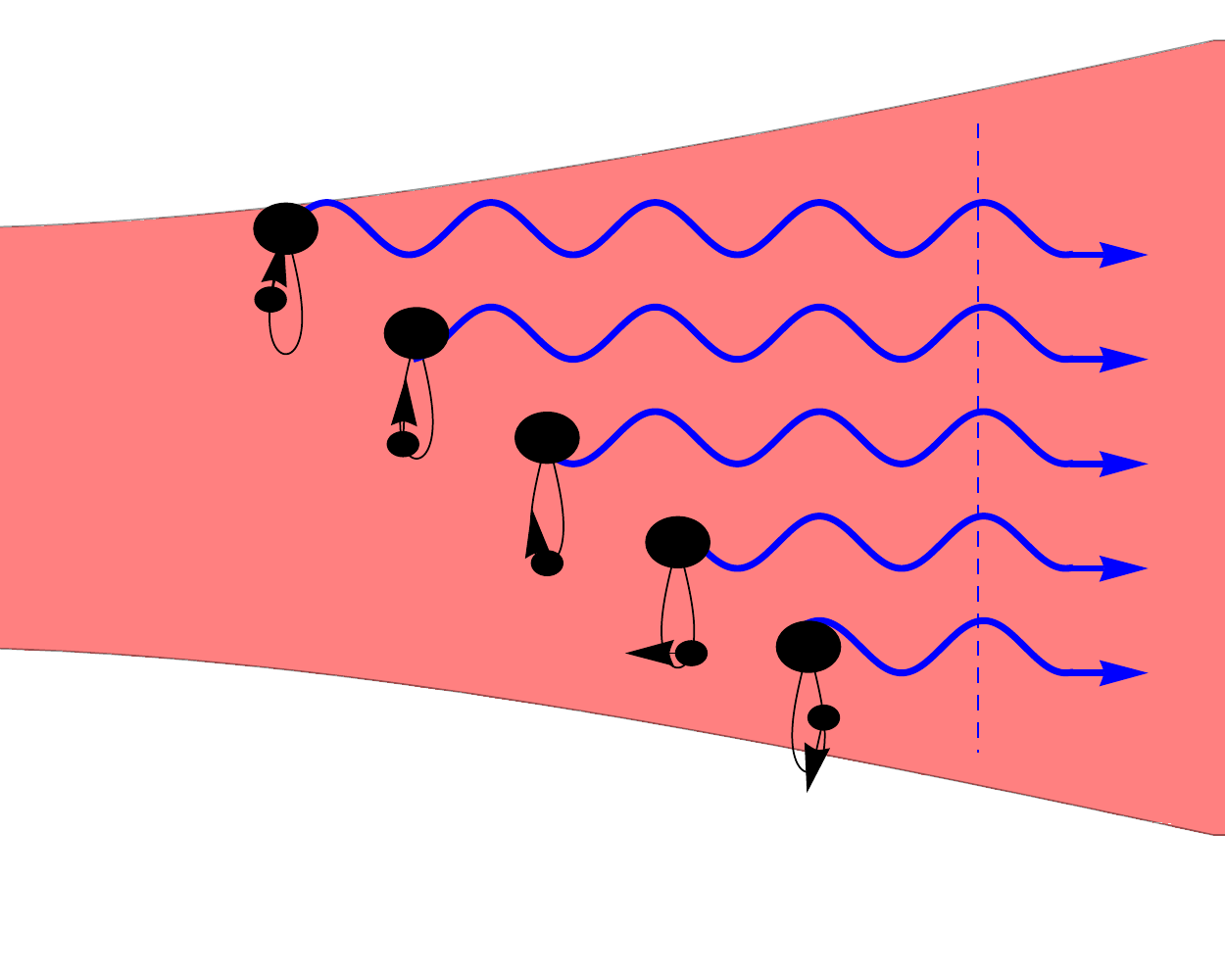}
\caption{\label{fig-collision}
(Color online) The situation of a coherent superposition of the emitted light
(blue wiggled line) from different atoms (black circles) is shown. The HHG
process of each atom in the Gaussian focus (red, gray area) is coherently
triggered by the laser field which propagates to the right.}
\end{figure}
For phase matching, essentially, the vectorial sum of the individual $n$
fundamental photon wavevectors has to equal the wavevector of the $n$th-harmonic
photon (i.e. photon momentum conservation) to ensure perfectly constructive
interference of all microscopic atomic emitters in the far field (see
Fig.~\ref{fig-collision}). For a plane-wave geometry, this would mean that the
total refractive index (governed by the neutral and the generated plasma
dispersion) of fundamental and high-harmonic frequencies has to be identical. 
In typical experiments, HHG is carried out near the focus (waist) of a laser or
in a low-diameter guided (wave guide, hollow fiber) geometry to realize high
intensities.  In this case, an additional geometric wavevector component,
particularly important for the long-wavelength fundamental beam, has to be taken
into account in phase matching. For a focusing beam, this arises from the
gradient of the Gouy Phase~\citep{GOUY1890,SIEGMAN1986} (for a Gaussian beam),
while for the guided geometry this contribution comes from the dispersion
equation of waveguide modes~\citep{MARCATILI1964} (physically from the fact that
the wavevector has a perpendicular component to create an effective radial node
at the waveguide wall).  Phase matching in focusing beams allows the analysis
and selection of individual microscopic quantum paths contributing to the high
harmonic spectrum~\citep{BALCOU:QP-99}.  In waveguides, fully phase-matched HHG
has been demonstrated in 1998~\citep{RUNDQUIST:PG-98,SCHNURER:GG-98}.  If
conventional phase matching is no longer possible, a spatial modulation with
periodicity $\Delta L$ of the target medium or its environment, resulting in an
additional effective wavevector $\Delta k=2 \pi/\Delta L$ to balance the momentum
conservation, can be employed to achieve quasi-phase
matching~\citep{PAUL:QP-03,BAHABAD:QP-10,Popmintchev:AN-10}.  Also without a
hollow-fiber guiding structure, self-guiding alone was shown to result in
efficient phase-matched HHG~\citep{TAMAKI:HE-99} with the potential for
attosecond pulsed emission~\citep{STEINGRUBE:DF-11}.  The high degree of spatial
coherence of the beam generated by a high-harmonic source has been demonstrated
experimentally by diffraction from spatially separated pinholes and other
objects~\citep{BARTELS:GS-02}. Below, in a), we will address the spectral
coherence measurements, leading to attosecond-pulsed characteristics, in more
detail.

Further major breakthroughs and disruptive technological advances on the HHG
source side included the reconstruction of the relative spectral phase of the
high-order harmonics to prove their attosecond structure~\citep{PAUL:OT-01}, the
mastery and control of the carrier-envelope phase (CEP) of the driving laser
pulses for isolated attosecond pulse generation~\citep{BALTUSKA:CE-03}, the spatial
and temporal control of phase matching
~\citep{PFEIFER:CS-05,PFEIFER:SC-05,WALTER:AF-06}, the gating
techniques~\citep{HENTSCHEL:AM-01,CORKUM:SF-94,SOLA:PG-06,Sansone:IS-06,
PFEIFER:SA-06,PFEIFER:HM-06,CHANG:DG-07,MASHIKO:DB-08,FENG:GI-09,PFEIFER:GB-07,
JULLIEN:IP-08,Abel:IG-99,FERRARI:AS-2010}), the comprehensive control over the
harmonic spectral shape towards attosecond pulse shaping~\citep{PFEIFER:CS-05},
the creation of high-energy pulses of HHG
light~\citep{HERGOTT:XU-02,MIDORIKAWA:XU-08,SKANTZAKIS:CR-09,TZALLAS:EU-11}, the
use of the shortest possible pulses of laser light, including the application of
light-field synthesis methods to essentially reach the half-optical-cycle
limit~\citep{WIRTH:ST-11}, and the long-wavelength drivers to enhance the
high-harmonic cutoff and reduce the
attosecond pulse chirp~\citep{DOUMY:IR-09,Popmintcheva:PM-09,SHINER:PC-11}. The
culmination of this progress nowadays allows the generation of sub-100-as
pulses, both isolated ~\citep{Goulielmakis:SC-08} as well as in pulse train
emission mode~\citep{KO:AS-2010}, and also reaching photon energies in excess of
1~keV~\citep{SERES:SC-05}.  Advanced methods to control and selectively generate
double or triple pulses---thus at the interface between isolated pulses and
pulse trains---are currently being explored both
theoretically~\citep{RAITH:TW-11} as well as
experimentally~\citep{PFEIFER:GB-07,MANSTEN:SS-09}, for applications in
attosecond interferometry. A few of these important milestones shall now be
described in somewhat more detail.

\emph{a) Demonstration of HHG spectral coherence and attosecond pulsed nature}\newline
Some time after the first observation of HHG~\citep{McPherson:SM-87,FERRAY:MH-88}, the question arose as to whether the individual separated harmonic "comb" lines were coherently locked in a fixed relative phase relation.  If their relative phases would be small on a scale of $\pi$, the consequence would be an attosecond pulse train.  The race was on to test this spectral phase relation.  As attosecond pulses typically come with low pulse energies, nonlinear-optical autocorrelation commonly applied for femtosecond pulses was not immediately possible, and still is not a viable option for all photon-energy ranges~\citep{TZALLAS:LB-03,SEKIKAWA:NL-04}.  Instead, the solution was a temporal cross correlation of the HHG light with a moderately intense and coherently locked copy of the 800~nm HHG driver pulse in a gas medium while observing photoelectron emission~\citep{PAUL:OT-01}.  In this process, the odd harmonic photons with energy $E_{2n+1}$ ionized an electron with excess energy $E_{2n+1}-I\I{p}$.
The additional presence of the 800~nm laser photons resulted in additional photoelectron energies at $E_{2n+1}-I\I{p}\pm \omega\I{L}$ by absorption or emission of 800-nm photons, the so-called sidebands at even-integer harmonic orders in between the peaks corresponding to the original odd harmonics.  Importantly, interference of the ionization pathways proceeding via two different harmonics $E_{2n+1}$ and $E_{2n-1}$ resulting in the same final electron energies $E_{2n+1}-I\I{p}-\omega\I{L}=E_{2n-1}-I\I{p}+\omega\I{L}$, allowed the detection of the relative phase between these harmonics from the constructive or destructive interference for the sideband peak amplitudes.  This method has been termed RABBITT (reconstruction of attosecond beating by interference of two-photon transitions) and is currently used, along with its complementary streak-field approach~\citep{HENTSCHEL:AM-01,Itatani:AS-02}, for a variety of scientific applications (see Section~\ref{EXP-APPLICATION}), including the measurement of photoionization dynamics.

\emph{b) Few-cycle and CEP control technology}\newline
After the successful demonstration of attosecond pulse trains, the next goal was to isolate one single attosecond pulse per driver pulse, in order to provide a temporally singular probe for quantum-dynamics measurements.  The group around Ferenc Krausz, being pioneers in the creation of the shortest optical laser pulses, were the first to measure, again by using photoelectron spectroscopy, an isolated attosecond pulse~\citep{HENTSCHEL:AM-01}.  It was soon realized that the thus-far uncontrollable CEP of the driving laser pulse played a paramount role in generating single vs.~double attosecond pulses per driver pulse (depending on whether the driver field within the envelope exhibited a cos-like or sin-line shape).   From a crucial cooperation with a pioneer in optical precision spectroscopy and creator of optical frequency comb technology, Theodor H\"ansch (physics Nobel prize in 2005~\citep{HAENSCH:NP-06}), the group was soon able to stabilize the CEP and thus the electric field within the envelope of the driver and selectively generate isolated attosecond pulses~\citep{BALTUSKA:CE-03}.  Applying the same methodology with further improvements of femtosecond laser and optics technology, this method currently allows the production of the shortest isolated attosecond pulses of ~80~as~\citep{Goulielmakis:SC-08}.

\emph{c) HHG gating techniques}\newline
As isolated attosecond pulses are desirable for many applications, early ideas focused on the isolation of attosecond pulses~\citep{CORKUM:SF-94} by means of employing the strong ellipticity dependence of the HHG process~\citep{BUDIL:IE-93}.  This dependence is immediately understood from the recollision model~\citep{Corkum:PP-93}, where only linearly polarized light can lead to the strict classical return of the electron to the atom, while for elliptically polarized light the electron would always miss its point-like origin.  In quantum reality, it is the delocalization of the electron wavefunction and the spatial extent of the atomic or molecular target system~\citep{FLETTNER:ED-02,FLETTNER:AM-03} that create a non-zero harmonic intensity response, which however remains a rapidly decaying function of ellipticity and typically reaches 10\% of its linear-polarization yield for ellipticities between 0.1 and 0.2.  A laser pulse exhibiting a temporally varying ellipticity was thus suggested~\citep{CORKUM:SF-94} and experimentally proven~\citep{SOLA:PG-06,Sansone:IS-06} to result in isolated attosecond pulses.  A major benefit of this method is the fact that the entire spectral bandwidth of an attosecond HHG pulse can be used for experiments, rather than only a spectral portion around its cutoff as is the case for linear-polarization methods~\citep{HENTSCHEL:AM-01}.  These latter methods are sometimes referred to as intensity gating based on the fact that attosecond pulse isolation depends on the rapid variation of intensity and thus HHG cutoff photon energy from one half-cycle to the next within the driving laser pulse.  The attosecond pulse has then to be selected by using a high-pass photon energy filter such as a metal foil before the experiment.  Other gating techniques were invented that rely on two-color driving methods, i.e. two-color gating~\citep{PFEIFER:SA-06,PFEIFER:HM-06,MERDJI:IA-07}, also in combination with a time-dependent polarization field, resulting in the so-called double-optical gating (DOG) method~\citep{CHANG:DG-07,MASHIKO:DB-08}, including a generalized (GDOG)~\citep{FENG:GI-09} way to use pulses deep in the multi-cycle regime for generating isolated pulses.  Yet another scheme suggests the gating of isolated attosecond pulses on the leading edge of the driver pulse by termination of HHG in a rapidly ionizing medium, either by the collective-medium response through phase matching control~\citep{PFEIFER:GB-07,JULLIEN:IP-08,Abel:IG-99} or by controlling the single-atom response by completely depleting the ground state of the atomic target system~\citep{FERRARI:AS-2010}.

\section{Hard x-ray HHG and zeptosecond pulses}
\label{sec:intro2}

The extension of HHG sources toward higher photon energies and shorter pulse durations is crucial to enable 
new applications  such as, e.g., time-resolved diffraction imaging~\citep{NEUTZE:PB-00} with sub-\AA{}ngstr\"om resolution, excitation of tightly bound core electrons
or even the time-resolved study of nuclear excitations~\citep{Burvenich:NQ-06,WEIDEMUELLER:NE-11}.
The advancements in the time and energy domain are inter-connected. A larger harmonic bandwidth is required to generate shorter light pulses. Presently, the keV regime of HHG is approached allowing to break the one attosecond barrier toward  pulses of zeptoseconds duration as visible from the time-bandwidth product [Eq.~\eqref{eq:time-bandw}].

\begin{figure}
\begin{center}
{\small (a)}
 \includegraphics[width=0.4\textwidth]{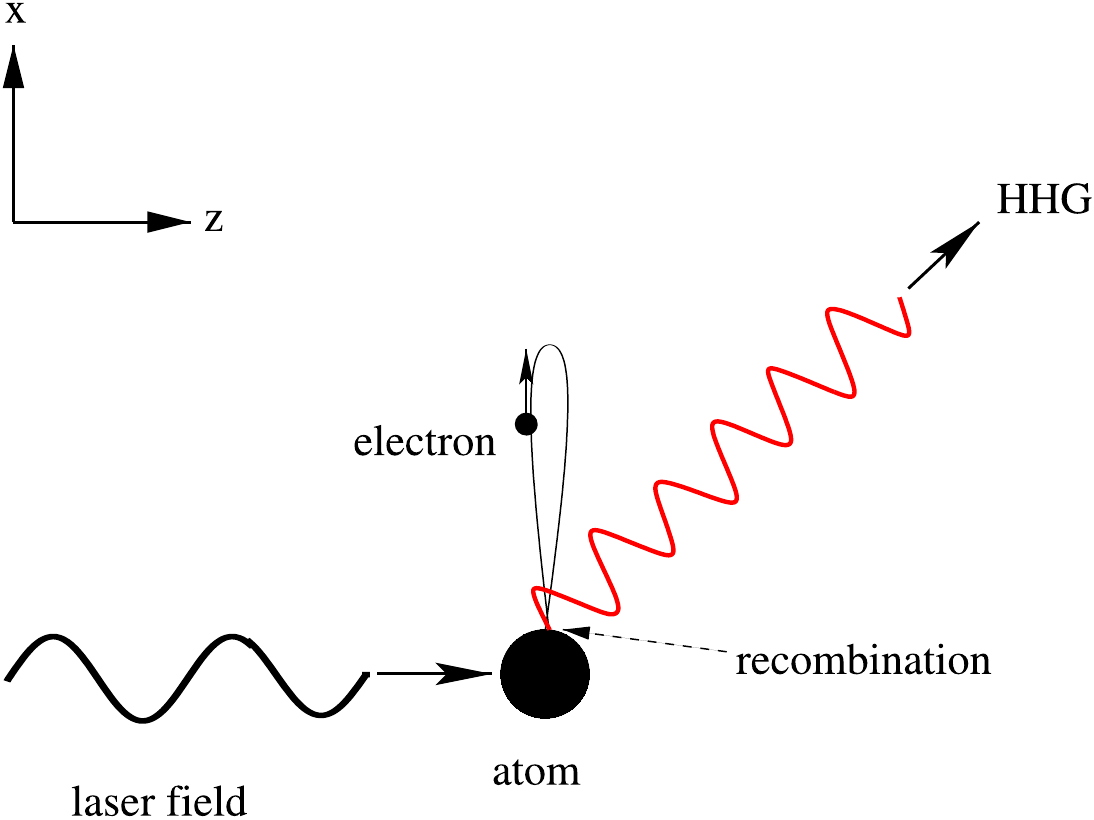}
        \hspace*{0.8cm}
{\small (b)}
 \includegraphics[width=0.4\textwidth]{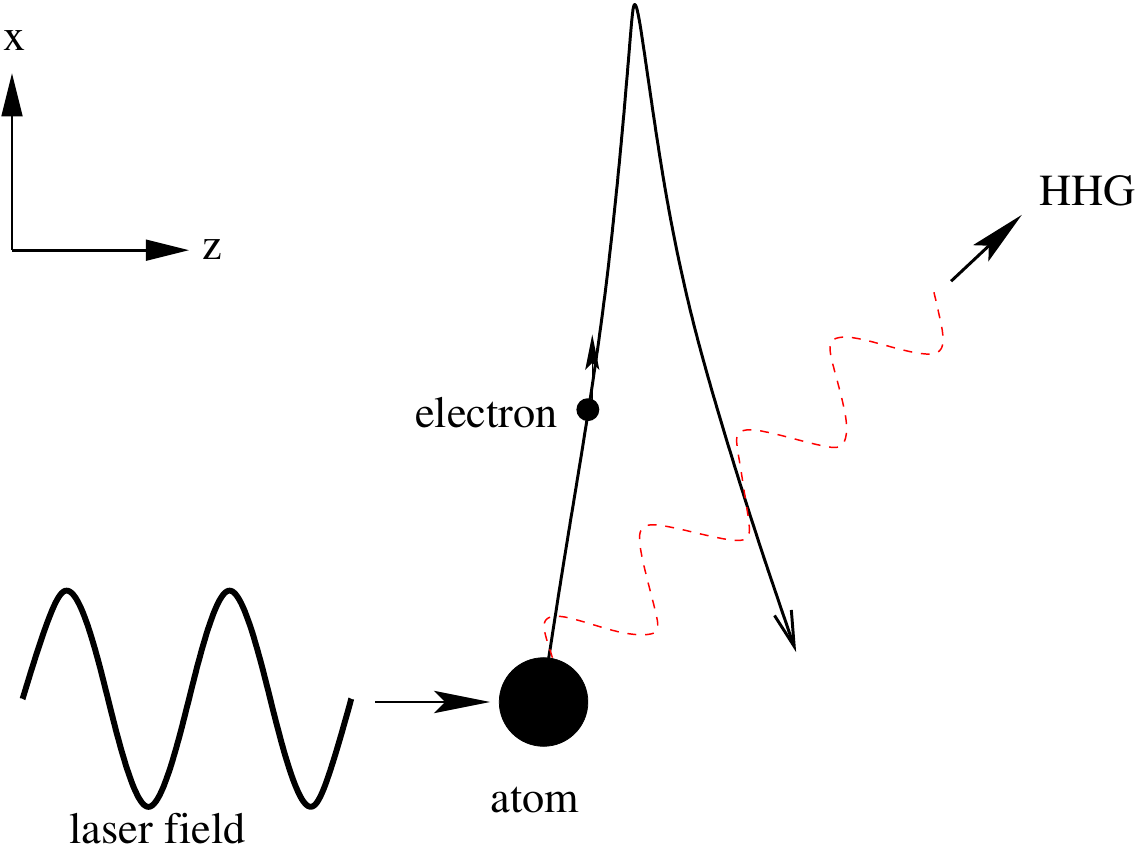}
\caption{(Color online) Two different classical electron trajectories for a linearly polarized laser field in the (a) nonrelativistic  and (b) highly relativistic regime. In the relativistic case the electron does not return to the atomic core when the laser field changes its sign. HHG is unlikely what is depicted by the dashed line. } \label{comp_traj}
\end{center}
\end{figure}

The spectral cutoff energy of HHG [see Eq.~\eqref{eq:cutoff}] can principally be increased by employment of higher laser intensities. This way, photon energies of few keVs have been generated~\citep{SERES:SC-05,Seres:XA-06} with small rates just above the detection threshold [see Fig.~\ref{fig-hhg-typical}~(b)] because this straightforward increase of the laser intensity is connected with several difficulties.
First, rising the laser intensity above resulting harmonic cutoff energies of few hundred electronvolts leads to a large free-electron background causing phase mismatch and tiny emission yields on a macroscopic level. For this reason, in a different approach~\citep{Popmintchev:EP-08,DOUMY:IR-09,Popmintcheva:PM-09,CHEN:LW-10,Ghimire:BC-11}, higher cutoff energies are obtained without significant change of the free-electron background  by increase of the driving laser wavelength for fixed  laser intensity. This way, increasing the ponderomotive potential without increasing intensity, one can potentially attain photon energies corresponding to the nonrelativistic HHG limit.
Second, relativistic effects can lead to a dramatic suppression of the single atom HHG yield~\citep{DiPiazza_2012}.
While the  nonrelativistic electron motion is mainly along the polarization direction of the laser field [see Fig~\ref{comp_traj}a], the Lorentz force becomes noticeable for electron velocities approaching the speed of light. Then  the electron is transfered along the propagation direction of the laser pulse hindering recombination and frustrating HHG, see. Fig~\ref{comp_traj}b. The relativistic drift becomes significant if the drift distance exceeds the wave packet dimensions after spreading. This 
happens at laser intensities of about $10^{17}\Wcm$~\citep{PALANIYAPPAN:ES-06} at $800\U{nm}$ wavelength and marks the limits in terms of photon energy for the nonrelativistic HHG at about $10$ keV.

For short pulse generation, besides a large bandwidth, synchronization of the emission of the different harmonic components in the pulse is required.
However, the HHG pulses have an intrinsic chirp, the so-called attochirp~\citep{MAIRESSE:AS-2003,DOUMY:IR-09} and, thus, are much longer than their bandwidth limit. 
To compress the emitted pulse down to its fundamental limit, dispersive elements of either chirped multilayer x-ray mirrors~\citep{MORLENS:CM-05},  thin metallic films~\citep{KIM:SS-04,LOPEZ:AP-05}, grating compressors~\citep{POLETTO:PC-08} or thick gaseous media~\citep{KO:AS-2010,KIM:SC-07} are employed.
However, with increasing bandwidths, it will be difficult to find media with suitable dispersion in the future.

In the following a selection of techniques for enhancing the cutoff energy or decreasing the pulse duration of the harmonic light are reviewed.

\subsection{HHG with long wavelength drivers}\label{sec:longwvl}

As indicated in the previous part, the quest of reaching keV photon energies and
zeptosecond pulses is ultimately linked with solving the phase-matching problem.
 For coherent growth of the harmonic signal, the phase velocities of the
harmonic and laser field are required to be equal. Under this condition, the light emitted
from different atoms adds up constructively as sketched in
Fig.~\ref{fig-collision} and explained in Sec.~\ref{EXP-SOURCE}.
The condition of equal phase velocities can also be defined as a vanishing phase-mismatch vector $\Delta k $~\citep{GAARDE:MA-08}.
The phase-mismatch vector is~\citep{DURFEE:PM-99,Popmintcheva:PM-09}
\begin{equation}
\label{eq:gen_phase-mismatch_gen}
 \Delta k\approx q k\I{geo}- q p (1-\eta) \frac{2\pi}{\lambda\I{L}}(\Delta \delta+n\I{2})+q p \eta N\I{a}r\I{e}\lambda\I{L}+\alpha \frac{\D I}{\D x}\, ,
\end{equation}
where $q$ is the harmonic order, $p$ the gas pressure, $\lambda\I{L}$ the laser wavelength, $\Delta \delta$ the difference of the index of refraction per pressure between the laser wavelength and the harmonic wavelength, $n\I{2}=\tilde{n\I{2}}I\I{L}$ the nonlinear index of refraction per pressure, $N\I{a}$ the number density of atoms per pressure, and $r\I{e}$ the classical electron radius. $q k\I{geo}$ is a geometric term
due to waveguide or near-focus propagation. The second and third term  cover the
refractive index difference between both waves caused by atomic and plasma
dispersion, respectively. The last term covers the variation of the atomic
dipole emission phase~\citep{LEWENSTEIN:PH-95} with the laser intensity
introduced in Eq.~\eqref{phase}. Moreover, non-adiabatic effects due to the
laser pulse deformation, which are important in the case of ultra-short driving
pulses especially close to the saturation regime, are excluded, see,
e.g.,~\cite{TEMPEA:SP-00,GEISSLER:SP-00}. For a detailed review on the
phase-matching aspects of HHG see, e.g.,~\cite{GAARDE:MA-08}.

In experiments, conditions have to be found to achieve vanishing $\Delta k$ [Eq.~\eqref{eq:gen_phase-mismatch_gen}]. We concentrate on the discussion of waveguide propagation where $\alpha \frac{\D I}{\D x}$ is negligible. It is beneficial that the two pressure-dependent terms of the atomic and plasma dispersion in Eq.~\eqref{eq:gen_phase-mismatch_gen} have opposite signs and, thus, can cancel each other to some extent for low ionization degrees  $\eta$.
Hence, phase-matching is commonly achieved by selecting the pressure $p$ such that $\Delta k$ vanishes exploiting that the pressure independent geometric term $q k\I{geo}$ balances the difference between the two dispersion terms that depends linear on $p$. 
At a fixed IR laser wavelength but increasing laser intensity, balancing is possible  to some extent to attain HHG energies up to about 100 eV depending on the atomic species.
For HHG cutoff energies above this limit,  the laser intensity is sufficiently high to ionize a considerable part $\eta$ of the atoms leading to a serious misbalance between the atomic and plasma dispersion.
Above this critical level of ionization $\eta\I{cr}$, it is not possible to counteract the misbalance by an increase of the gas pressure any more.
An obvious approach to circumvent this problem is to keep the laser intensity on a low level but to increase the driving wavelength. This way, the quadratic scaling of the single-atom cutoff energy [Eq.~\eqref{eq:cutoff}] with the driving wavelength is exploited while the ionization level $\eta$ remains constant because the ionization rate in the tunneling regime depends mainly on the laser intensity rather than its wavelength. Long time, the approach was not pursued because the single-atom yield has an unfavorable scaling with the driving wavelength of $\sim\lambda^{-5.5\pm0.5}$~\citep{Corkum:PP-93,GORDON:SH-05,Tate:SD-07,Colosimo:ST-08,Frolov:WS-08,DOUMY:IR-09,Schiessl:WS-07,Hernandez:sw-09,Shiner:WS-09}. Despite this, the generation of bright soft x-ray harmonics with mid-infrared drivers has been demonstrated recently~\citep{Popmintchev:EP-08,Popmintcheva:PM-09,CHEN:LW-10} 
exploiting the fact that near the critical level of ionization, high gas pressures are required to achieve phase-matching. For high gas pressures, the 
increased number of emitters compensates the decrease of the single-atom response yielding  efficient overall HHG.
Gas-filled capillaries are especially suitable in this regard~\citep{DURFEE:PM-99}. This way efficient generation of HHG with several hundreds of electronvolts bandwidth has been demonstrated~\citep{CHEN:LW-10} being sufficient to support pulses of only $11\U{as}$ duration.
From this point of view, the concept of increasing the driver wavelength seems highly promising at the moment.

In the case that the phase-matching condition $\Delta k\approx0$ is not feasible, 
so-called quasi-phase-matching (QPM) schemes are invoked. Their principle relies on symmetry breaking between the positive and negative contributions responsible for the destructive interference. In this regard a weak counterpropagating IR field~\citep{PEATROSS:SZ-97,COHEN:GA-99}, weak static fields~\citep{SERRAT:SE-10}, modulated wave guides~\citep{CHRISTOV:MG-00,PAUL:QP-03,GIBSON:CS-03} or multiple gas jets~\citep{Wilner:QP-11} can be employed. However, these schemes require additional efforts in the experimental implementation.

\subsection{Relativistic regime of HHG}\label{sec:drift}

Although the approach in Sec.~\ref{sec:longwvl} to increase the driver wavelength raises the perspective to construct bright HHG sources above one keV, limits will appear in the multi-keV regime for several reasons. First, the velocity of the electron will become comparable to the speed of light and, thus, the electron will undergo a drift motion caused by the Lorentz force preventing recollision along with recombination [see also the discussion in the beginning of Sec.~\ref{sec:intro2}]. Second, the increase of the driver wavelength is accompanied by a longer electron excursion which can exceed the mean distance between the atoms at the required high gas pressures. The resulting collisions of the electron with other atoms 
are a source of incoherence and prevent efficient HHG~\citep{STRELKOV:HD-05}. Facing these difficulties, strategies are required to overcome the relativistic drift and to solve the phase-matching problem in a gas of multiply-charged ions rather than using long-wavelength drivers on a low ionization level.

While in the weakly relativistic regime the drift problem is not severe and harmonic emission can be observed~\citep{LATINNE:RH-94,HU:DM-01,Keitel:CX-02}, at higher intensities special methods have to be applied to counteract the relativistic drift.
To suppress the drift, highly charged ions moving relativistically against the laser propagation direction~\citep{MOCKEN:BA-04,CHIRILA:IS-04}, antisymmetric molecular orbitals~\citep{Fischer:AM-06} or a gas of positronium atoms~\citep{HENRICH:PI-04,HATSAGORTSYAN:ML-06} can be used. Different combinations of laser fields  have also been proposed for this purpose such as a tightly focused laser beam~\citep{Lin:HH-06}, two counterpropagating laser beams with linear polarization~\citep{KEITEL:RH-93,KYLSTRA:BS-00,TARANUKHIN:RH-00,VERSCHL:SW-07} or with equal-handed circular polarization~\citep{MILOSEVIC:DA-04,VERSCHL:SW-07}. In the latter field configuration, the relativistic drift is eliminated. However, in this scheme the phase-matching is particularly problematic to realize~\citep{LIU:LG-09}. In the weakly relativistic regime, the Lorentz force can also be compensated by a second weak laser beam being polarized in the strong beam propagation direction~\citep{CHIRILA:ND-02}. Two consecutive laser pulses  or the laser field assisted by a strong magnetic field have been proposed as well~\citep{VERSCHL:RF-07,VERSCHL:RC-07} 
but this requires large magnetic fields and dilute samples. 
In~\cite{KLAIBER:TA-06,KLAIBER:LO-07}  it has been shown that the relativistic drift can be significantly reduced by means of a special tailoring of the driving laser pulses in the form of attosecond pulse trains (APTs). 
These results concern  the relativistic drift problem only. However, in order to generate relativistic harmonics, both the drift and the phase-matching problems have to be tackled simultaneously. Below we report on two works~\citep{KOHLER:HX-11,Kohler:Ma-12} approaching those jointly.

\begin{figure}[t]
\begin{center}
 \includegraphics[width=0.9\textwidth]{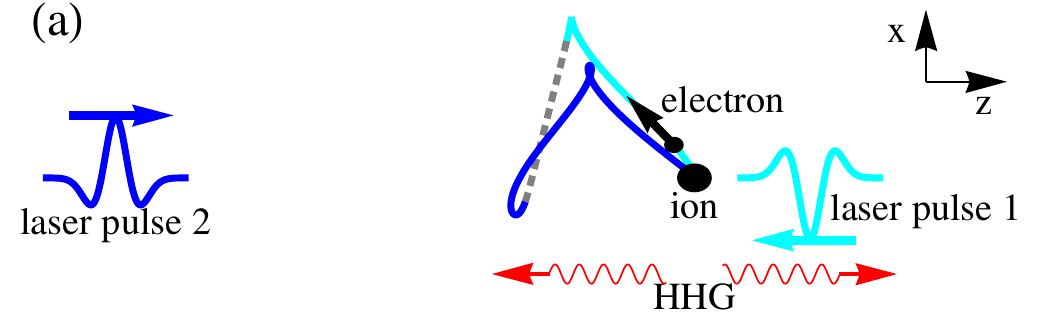}\\
\vskip 0.3cm
\includegraphics[width=0.55\textwidth]{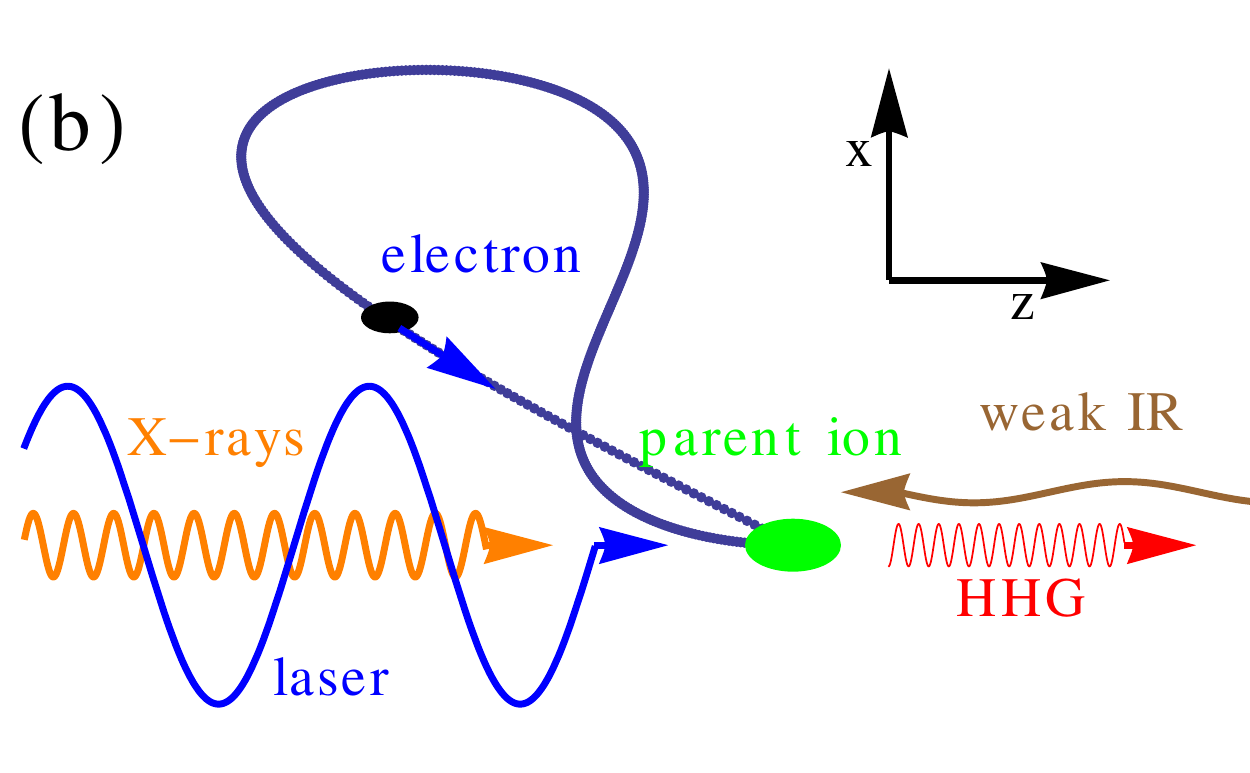}\\
\caption{(Color online) (a) Single-atom perspective of the HHG setup with
counterpropagating APTs. The classical trajectory of a rescattered electron of a
single atom in the gas target. After ionization by pulse 1, the ejected electron
is driven in the same pulse (light blue, light gray), propagates freely after
the pulse has
left (gray dashed) and is driven back to the ion by the second laser pulse (dark
blue, dark gray). Figure reprinted with permission from~\cite{KOHLER:HX-11}. 
Copyright 2011 by the
Institute of Physics.
(b) Geometry of the HHG process for a collinear alignment of the x-ray and laser
fields. A weak IR field  is added that accomplishes phase-matching.
}
\label{cprop_setup_a}
\end{center}
\end{figure}

In~\cite{KOHLER:HX-11}, HHG emission from a gaseous medium driven by two counterpropagating APTs has been investigated. The field configuration is capable of circumventing the impact of the relativistic drift~\citep{KLAIBER:ZS-07,HATSAGORTSYAN:UFO-08} as seen from the schematic of the electron dynamics in Fig.~\ref{cprop_setup_a} (a).
The electron is liberated by laser pulse 1 which reaches the atom first. Subsequently, it is driven by this pulse in the continuum and undergoes the relativistic drift. This part of the trajectory is indicated by the light blue coloring [in Fig.~\ref{cprop_setup_a} (a)]. Thereafter, the electron propagates freely (gray dashed) in the continuum, before, a moment later, the second pulse reaches the electron, reverts the drift and imposes recollision (dark blue). 
This setup has a distinguished property which facilitates the realization of phase matching despite the significant free-electron dispersion. In fact, for different atoms situated along the propagation direction, the time delays between the first pulse and the second pulse are different. This is reflected in different intrinsic harmonic emission phases.
 The phase can be tuned by variation of the laser field intensity to compensate the phase mismatch caused by the free electron background. The period of the spatial modulations is given by the APT period.  

In~\cite{KLAIBER:CH-08,HATSAGORTSYAN:UFO-08,Kohler:Ma-12} relativistic HHG is achieved employing a setup where the laser field is assisted by an x-ray field. 
The usefulness and applicability of x-ray assistance have been proved in various theoretical investigations and experiments for different purposes (see Sec.~\ref{sec:xuv-ass}). 
In the present case, the assisting x-ray field is required for the ionization
via one x-ray photon absorption. Therefore,
the x-ray frequency has to exceed the ionization energy conveying the electron a large enough initial momentum opposite to the laser propagation direction 
which is able to compensate for the relativistic drift. 
The setup is displayed in Fig.~\ref{cprop_setup_a}~(b). A collinear alignment of the laser and x-ray field propagation directions is shown to be advantageous in terms of phase-matching~\citep{Kohler:Ma-12}. An additional weak counterpropagating IR field arranges a quasi-phase-matching scheme.

Both of the former relativistic HHG  setups provide a tiny macroscopic HHG yield of $10^{-7}$ photons per shot for a HHG energy of about $50\U{keV}$ in common. Several issues have been identified lowering the HHG yield when extending the HHG cutoff towards the multi keV regime~\citep{KOHLER:HX-11,Kohler:Ma-12}. To discuss them, the macroscopic yield can be approximately expressed by the product
\begin{equation}
N=\frac{\D w_n}{\D \Omega} \times \Delta n \times \Delta\Omega \times \Delta t  \times V^2 \rho^2 ,\label{estimation}
\end{equation}
where  $\D w_n/\D \Omega$ is the single-atom emission rate, $\Delta n$ is the number of harmonics within the phase-matched frequency bandwidth, $\Delta t$ the interaction time that is approximately the delay between both pulses, $\Delta \Omega$ the solid angle of emitted harmonics,
$V$ the volume of coherently emitting atoms (perfect phase-matching is assumed in this volume) and $\rho$ the atomic density. 
The terms in Eq.~\eqref{estimation} can be evaluated for the two relativistic setups as well as for an experiment with standard HHG in the nonrelativistic regime, see e.g.,~\cite{Goulielmakis:SC-08}. Comparing the factors in Eq.~\eqref{estimation} between the different setups, one notes that in the relativistic regime suppression arises from the single-atom yield, the smaller solid emission angle and the lower gas density. The reasons for the lower single-atom yield are manifold: The higher momentum of the recolliding electron favors scattering instead of recombination. Furthermore, the larger energy range of the recolliding wave packet leads to a smaller HHG yield per harmonic. Then compared to a non-relativistic sinusoidal field only, the ionization and continuum dynamics of the wave packet are changed and a smaller fraction actually revisits the core area. In analogy to the interference pattern of an aperture, the solid emission angle decreases quadratically with the wavelength. Additionally, phase-matching could only be accomplished for either low density ($10^{16}$ cm$^{-3}$ in \citet{Kohler:Ma-12}) or small volume (length $\sim 12.5\,\mu$m in~\citet{KOHLER:HX-11}). Note that the formerly discussed reasons of surpression are not caused by relativistic effects rather than by the high HHG energy in general.

On the bottom line, the former results show that more efforts are required to improve the efficiency of HHG in the multi keV regime. 
Below we will discuss another way to increase the HHG energy without increasing the laser intensity and suffering from drift problems.

\subsection{XUV-assisted HHG}\label{sec:xuv-ass}

The usefulness of XUV light assisting a strong laser field has been demonstrated for various purposes. 
It has been used to enhance HHG by many orders of magnitude compared with the case of the laser field alone~\citep{ISHIKAWA:PI-03,TAKAHASHI:DE-07}.
When the XUV field has the form of an attosecond pulse train  a single quantum path can be selected to contribute to HHG  and in this way allows to manipulate the time-frequency properties of the harmonics as well as to enhance a selected bandwidth of harmonics~\citep{SCHAFER:PC-04,GAARDE:DE-05,FARIA:HO-07}, and to extend the spectrum~\citep{FLEISCHER:GT-08R}.

Tuning an intense XUV field from an FEL to a resonance between a core and valence state can lead to the emergence of a second plateau that is shifted to higher energies by the former resonance energy with respect to the first plateau~\citep{Buth:HO-up}. The presence of originally two bound electrons is thereby crucial for the effect. Plateau extension in the presence of two electrons has already been noticed in \cite{KOVAL:DR-07}, however, with a low probability of the secondary plateau.  In \cite{Buth:HO-up}, the intensity of the secondary plateau is tunable via the FEL intensity.
\begin{figure*}[ht]
\begin{center}
\includegraphics[width=0.9\textwidth]{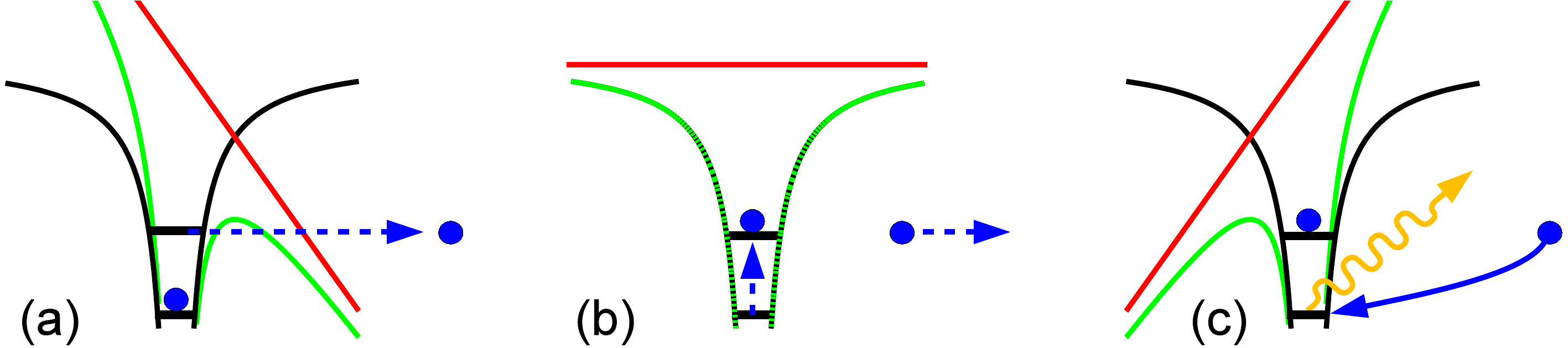}
\caption{(Color online) Schematic of the HHG scenario as a three-step process: (a) the valence electron is tunnel ionized; (b) the additional high-frequency light excites the core electron; (c) the continuum electron recombines with the core hole. Figure reprinted with permission from~\cite{Buth:HO-up}.  Copyright 2011 by the
Optical Society of America.
} \label{fig:rabi-schematic}
\end{center}
\end{figure*}
A schematic of the proposed scheme is shown in Fig.~\ref{fig:rabi-schematic}. The atoms are irradiated by both an intense optical laser field and the resonant x-ray field from an FEL.  
As soon as the valence electron is tunnel ionized by the optical laser field, the core electron can be excited to the valence vacancy by the x rays. Then the continuum electron, returning after a typical time of 1 fs, can recombine with a core hole rather than with the valence hole from that it was previously tunnel ionized and thus emit a much higher energy.

\begin{figure}[h]
\begin{center}
 \includegraphics[width=0.49\textwidth]{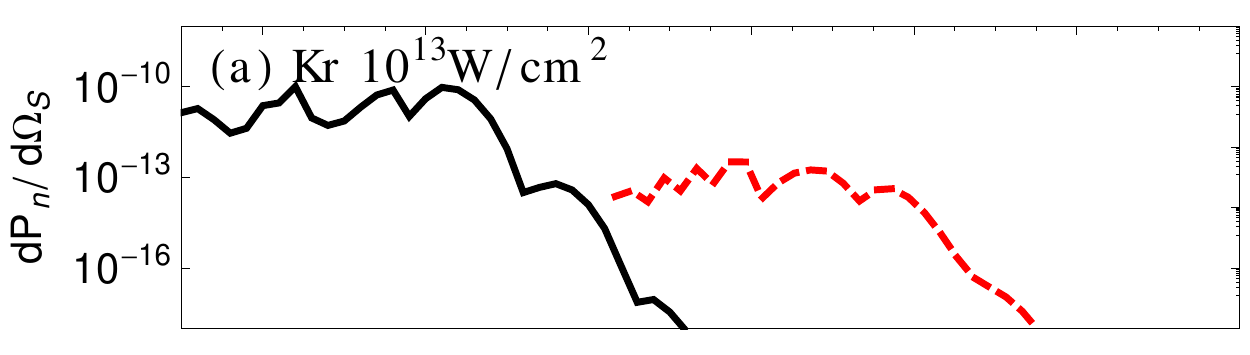}\hskip 0.2cm
 \includegraphics[width=0.49\textwidth]{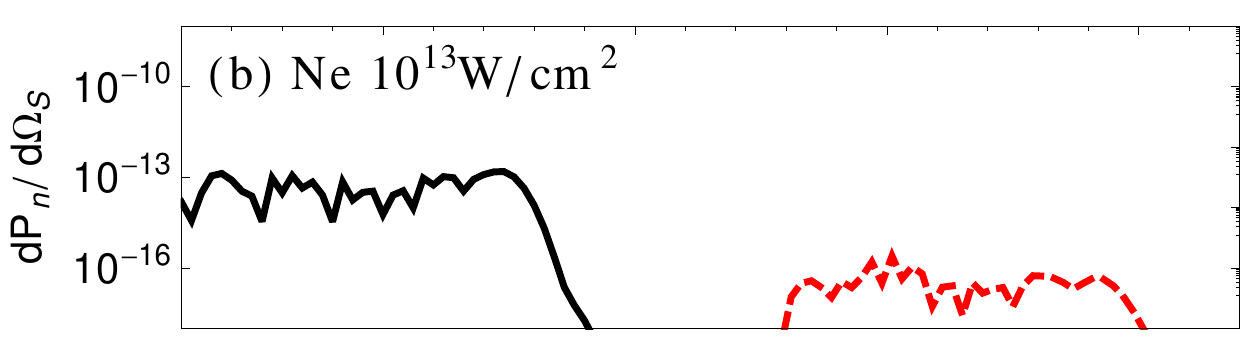}\\
 \includegraphics[width=0.49\textwidth]{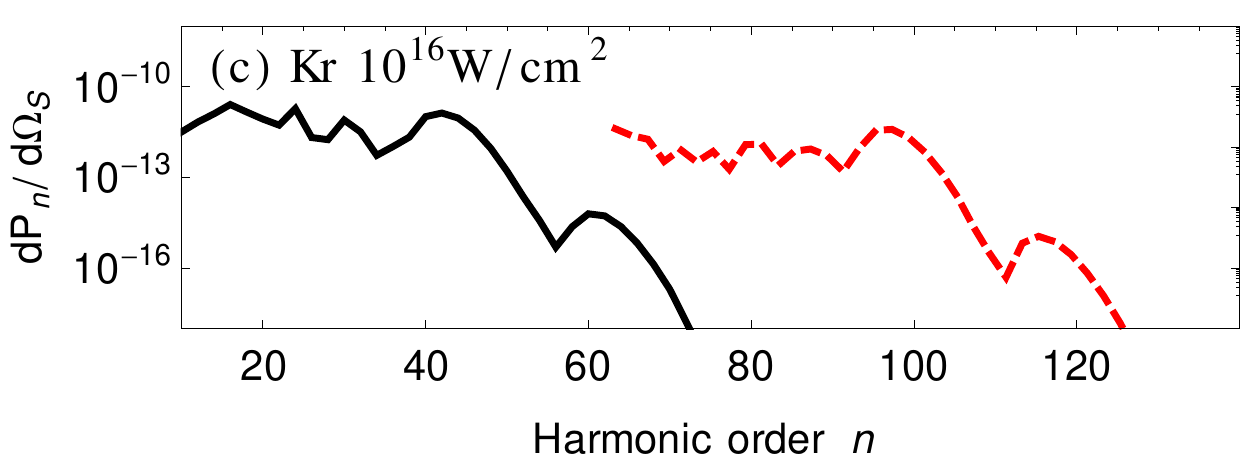}\hskip 0.2cm
 \includegraphics[width=0.49\textwidth]{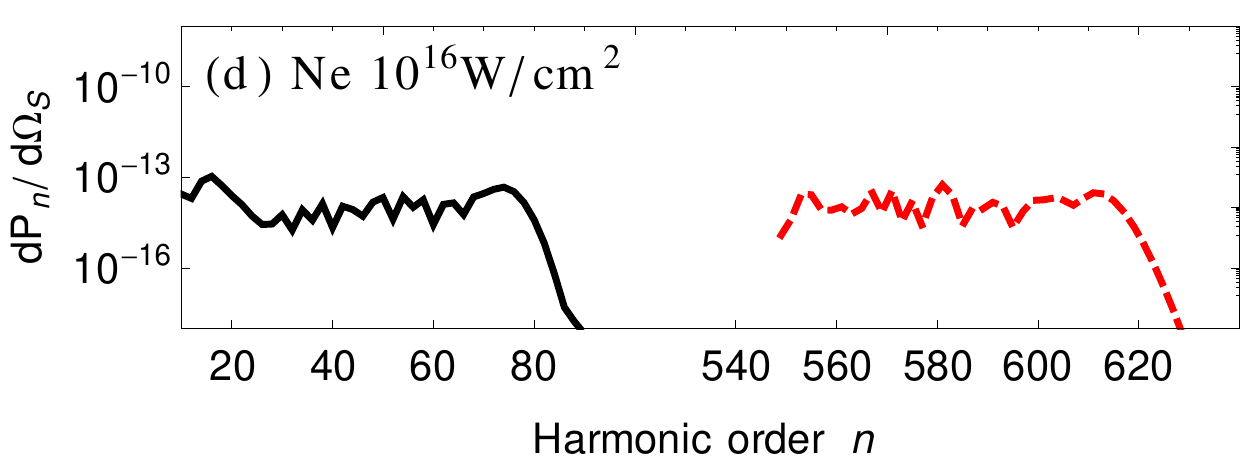}
\caption{(Color online) HHG photon numbers of the $n$-th harmonic for different 
x-ray intensities from (a), (c) krypton and  (b), (d) neon. The solid black line
stands for the recombination to the valence state whereas the more energetic
plateau arising from core hole recombination is the red dashed line.  The XUV
and laser pulse durations  are three optical laser cycles in all cases. Figure
adapted from \cite{KOHLER:IC-12}.} \label{fig:Rabi-cont}
\end{center}
\end{figure}

In~\cite{Buth:HO-up,Buth:HO-PA} an analytical formalism is developed to cope with the two-electron two-color problem. Losses due to tunnel ionization and direct x-ray ionization are included via  phenomenological decay constants in conjunction with Auger decay of the intermediate hole. 
The theory is applied to the $3d \rightarrow 4p$ resonance in a krypton cation as well as to the $1s \rightarrow  2p$ resonance in a neon cation. The results for a resonant sinusoidal  x-ray field for two different intensities are shown in Fig.~\ref{fig:Rabi-cont}. The chosen optical laser field intensity is $3\times10^{14}\Wcm$ for krypton and  $5\times10^{14}\Wcm$ for neon both at $800\U{nm}$ wavelength.

The most striking feature in the obtained spectra is the appearance of a second plateau. It is up-shifted in energy with respect to the first plateau by the energy difference between the two involved core and valence states. The two plateaus have comparable harmonic yields for x-ray intensities above $10^{16}\Wcm$. Quite importantly, the losses due to x-ray ionization do not lead to a significant drop of the HHG rate. The second plateau bears signatures of the core state and may offer a route for ultrafast time-dependent chemical imaging of inner shells~\citep{Itatani:TO-04,MORISHITA:MT-08}. Moreover, by exploiting the upshift in energy, attosecond x-ray pulses come into reach.

\subsection{Exotic light sources}

A route to shift the attainable cutoff energy to the multi-keV regime without suffering from the relativistic drift is the use of muons rather than electrons bound to the atomic system. According to the higher mass of the muon $m\I{\mu}\approx207$ compared to the electron mass $m\I{e}=1$, its kinetic energy is increased by a factor of the reduced mass $m\I{r}=m\I{\mu}m\I{n}/(m\I{\mu}+m\I{n})$ for the same velocity where  $m\I{n}$ is the nuclear mass. Thus, the attained energy of a muon is much higher for nonrelativistic velocities than for electronic systems. This can be expressed more quantitatively by the linear scaling of the ponderomotive energy $U\I{p}=\frac{m\I{r} \xi\I{rel}^2}{4}$ at the non-relativistic limit defined by the field strength parameter $\xi\I{rel}$ which has a weak dependence on $m\I{r}$~\citep{HATSAGORTSYAN:UFO-08}.
Note that higher laser intensities are required to accelerate the muon to the non-relativistic limit as can be seen from $\xi=\frac{E}{m\I{r} c \omega\I{L}}$. The high recollision energies could allow to trigger nuclear reactions~\citep{Chelkowski:MM-04}, to probe the nuclear structure due to the smaller Bohr radius of the bound muon~\citep{Shahbaz:NS-07} and to create zeptosecond pulses~\citep{Xiang:ZS-10}.  A precaution for this kind of experiments is certainly the realization of a target density being sufficiently high to generate a measurable emission yield.

Based on a different approach, \cite{Ipp:YS-09} proposed a method devised to generate yoctosecond pulses. The authors show that quark gluon plasmas generated by high-energy heavy ion collision can be a source of flashes of GeV $\gamma$-rays having such a short duration. Thereby, the fast expansion of the quark gluon plasma after collision is responsible for the short time scale. Under certain conditions a momentum anisotropy of the expanding plasma may form at intermediate times which leads to a preferential photon emission direction perpendicular to the collision axis during that short time. Therefore, a double pulse structure is emitted then in the other directions which has a principally variable pulse delay. The scheme could be employed for pump--probe measurements on the yoctosecond time scale.
\section{HHG in shaped driving pulses}
Femtosecond pulse shaping of the driving laser field~\citep{Weiner:FS-90,Zeek:FS-99,Kornaszewski:FS-08} is one of the most direct ways for controlling HHG. It allows for the precise adjustment of the ionization rates and the control of the electron continuum dynamics and, thus, for almost arbitrarily engineering the recolliding electronic wave packet with impact on the spectral and temporal properties of the emitted harmonic light.

\subsection{HHG yield and cutoff enhancement}

First experimental results using shaped driver laser pulses have demonstrated the enhancement of HHG, including the selective enhancement of individual harmonic orders~\citep{Bartels:SP-00}.  Using similar methods, control over the spectral positions of the produced harmonic spectral comb lines has also been demonstrated~\citep{Reitze:EH-04}.  

\begin{figure}[h]
\begin{center}
 \includegraphics[width=0.6\textwidth]{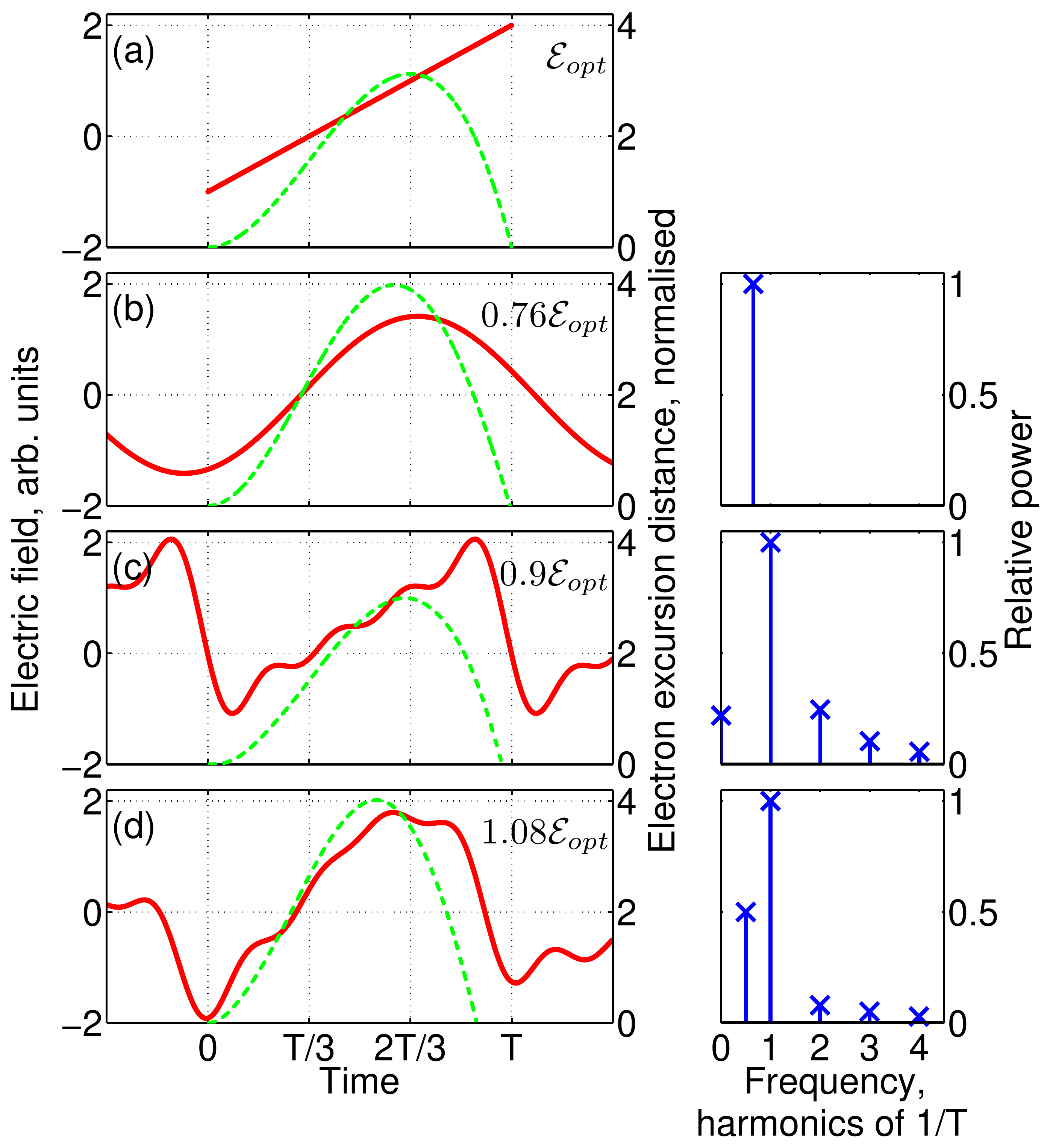}
\caption{(Color online) The solid red lines represent the electric field 
whereas the green dashed line is the respective electron excursion. (a) Optimal
waveform in terms of recollision energy. (b) Sinusoidal waveform similar to the
form in (a) but with a longer period ($1.54 T$). (c) Waveform of (a) composed of
the first four Fourier components. (d) As (c), but  optimized with respect to
the ionization rate. Figure reprinted with permission
from~\cite{CHIPPERFIELD:IW-09}. Copyright 2009 by the American Physical
Society.} \label{fig:chipperfield}
\end{center}
\end{figure}

On the theoretical side, \cite{CHIPPERFIELD:IW-09} have recently proposed an optimal pulse shape for reaching the maximum cutoff under a given laser pulse energy and laser period. The saw-tooth-form with a DC offset (red) in Fig.~\ref{fig:chipperfield}a turned out to be optimal~\citep{RADNOR:CS-08,Chipperfield:PW-10} with a cutoff energy about 3 times higher than a sinusoidal pulse of the same energy. The respective classical trajectory leading to the cutoff energy is shown as green dashed line. The optimal waveform has its highest intensity at the end of the cycle because this way it nurtures the photon energy the most just before recollision. The electric field in the initial part of the trajectory is only required to drive the electron away. High intensities along with high velocities in this part of the trajectory are not required for reaching the maximum cutoff but are at the expense of the pulse energy and, thus, would reduce the electric field required for the final acceleration. For this reason the intensity is lower in the beginning of the trajectory than at the end.
In Fig.~\ref{fig:chipperfield}c the authors demonstrate that the first four Fourier components are  sufficient to reproduce the optimal waveform to an adequate level. However, the pulse form lacks in efficiency with regard to the generated harmonics because the initial electric field is only about half of the final electric field. Thus, depending on the ionization potential the ionization rate is either small or over-the-barrier ionization happens at the end of the pulse suppressing HHG (see Sec.~\ref{sec:intf-model}). For this reason, the field is modified in the beginning of the period by a field of doubled wavelength to enhance the ionization rate. Quantum simulations demonstrate a similar efficiency for the generated harmonics compared to the case of a sinusoidal field.

Pulse shaping can also be employed to suppress the relativistic drift. The classical trajectory of a recolliding electron can be divided into a part where the electron moves away from the core and another where it returns to the core. In both parts of the pulse, the electron undergoes a relativistic drift motion. To reach the final high relativistic recollision energy the electron is not required to move during the first part of the trajectory with relativistic velocities  in the strong field. For that reason the pulse is shaped in~\cite{KLAIBER:TA-06} such that the field is negligible during the first part of the trajectory and, thus, the drift is avoided then.

\subsection{Attosecond pulse shaping}
Full control over attosecond pulse shapes however requires the comprehensive control over the HHG spectrum in both its amplitude and phase.  First experiments along these lines have focused on the control of the broadband harmonic spectral shape~\citep{PFEIFER:CS-05}, suggesting a collective medium response (phase-matching) mechanism for the observed controllability~\citep{PFEIFER:SC-05,WALTER:AF-06}.  Groups of harmonics as well as various individual harmonics could be selectively generated.  Also suppression of a single harmonic peak in an extended spectrum was possible.  The former spectral-amplitude shaping capability alone can result in major modifications of the attosecond pulse shapes that are produced.  An experiment applying such variably shaped light fields to an SF$_6$ molecular target system in an approach towards coherent electron control with shaped attosecond pulses demonstrated selectivity in the branching ratio of dissociative photoionization channels~\citep{PFEIFER:TO-07}.  Recent advances in the application of pulse shaping to HHG have also been reviewed in~\cite{Winterfeldt:OC-08}.

The former paragraph mainly dealt with the engineering of the spectral HHG amplitude.
 Phase control of high-harmonic spectra has been studied by means of passive dispersive effects~\citep{LOPEZ:AP-05,STRASSER:CI-06} or by active laser control of molecular alignment~\citep{Boutu:CA-08}.  A multi-dimensional control scheme including the CEP and broadband spectral driver pulses in the few-cycle regime for enhanced attosecond pulse control of amplitude and phase has also recently been suggested~\citep{RAITH:TW-11}.
Once available, fully controlled and intense attosecond fields may enable applications such as the preparation of exotic atomic and molecular electronic states, possibly leading to the creation of novel types of bonding which is out of reach of traditional (thermodynamics) methods.  It may also provide access to an unimaginable technology in molecular electronics, where (possibly quantum) information is created upon, processed and computed while traveling along and crossing tiny molecular wires exhibiting conjugated electronic bonds and the correspondingly delocalized electronic wavefunctions. 

Nowadays pulses down to a duration of 63~as~\citep{KO:AS-2010} have been generated and the bandwidth to create pulses of only 11~as is available~\citep{CHEN:LW-10}. 
Although the emitted harmonic light possesses a large bandwidth its time structure does not automatically correspond to its Fourier limit [see Eq.~\eqref{eq:time-bandw}] because of the attochirp stemming from the HHG process.
To compress the emitted pulse down to its fundamental limit, dispersive elements are employed as mentioned earlier. However, it may become demanding in the future to find suitable elements for multi keV bandwidths. An alternative way to circumvent the attochirp problem would be to modify the HHG process such that the light is emitted without attochirp. 
To achieve this goal, the continuum dynamics can be altered by adding a second-harmonic~\citep{ZHENG:DC-09} or a subharmonic~\citep{ZOU:DC-10} to the laser field. Significant chirp compensation has been proved in this case.

A method to acquire a complete control of the chirp has been proposed in~\cite{KOHLER:CF-11} 
by means of laser pulse shaping and soft x-ray assistance using an ionic gas medium.
This method allows for the formation of attosecond pulses with arbitrary chirp, including the possibility of attochirp-free HHG and bandwidth-limited attosecond pulses. 
\begin{figure}
\begin{center}
\includegraphics[width=0.7\textwidth]{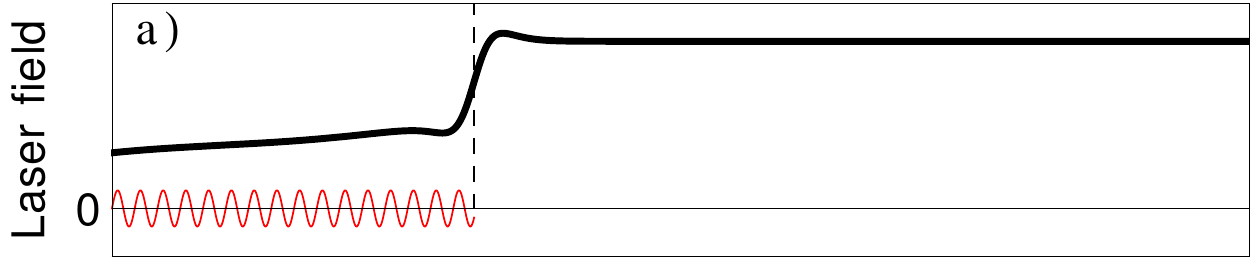}\\
\hskip0.0cm\includegraphics[width=0.7\textwidth]{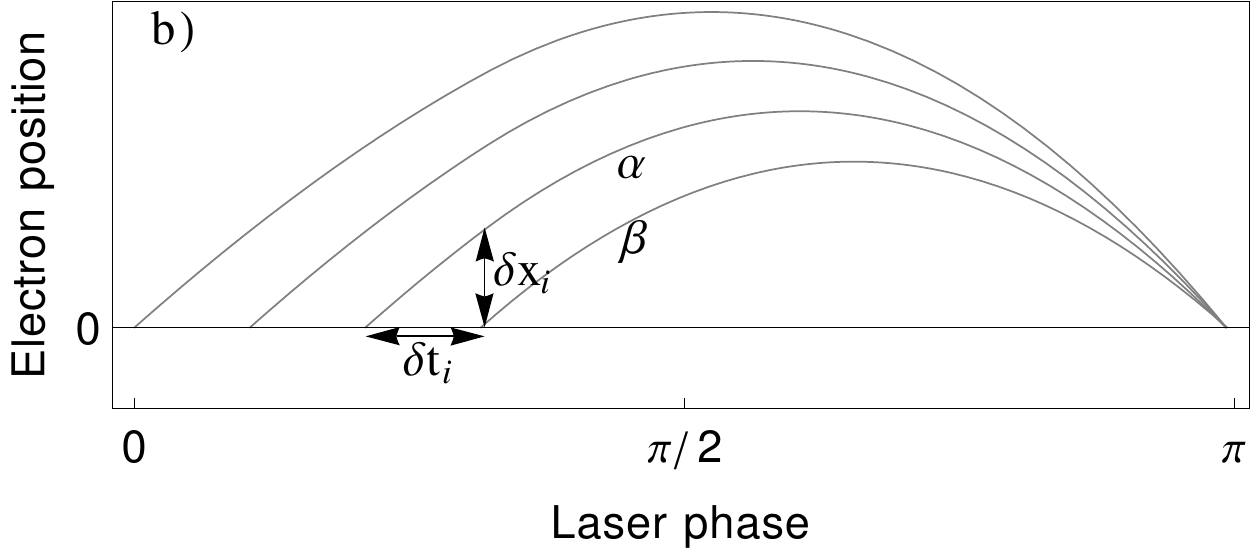}
\caption{(Color online) Schematic of the recollision scenario: a) A half cycle
of the tailored laser field (black). The wiggled red (gray) line is the field
of the assisting x-ray pulse. b) Different one-dimensional classical
trajectories in the field of (a) which start into the continuum at different
times but revisit the ionic core at the same time. Figure reprinted with
permission from~\cite{KOHLER:CF-11}. Copyright 2011 by the Optical Society of
America.} \label{figopti}
\end{center}
\end{figure}
The principle is illustrated in Fig.~\ref{figopti} which shows the relevant trajectories in the shaped field providing HHG. Since the recollision time of a certain harmonic can be identified with its group delay~\citep{MAIRESSE:AS-2003,KAZAMIAS:IC-04} in the emitted pulse, a simultaneous recollision of all trajectories leads to a bandwidth-limited attosecond pulse. The demand of simultaneous recollision can be fulfilled if the electron is freed by single-photon ionization when the x-ray frequency $\omega\I{X}$ is much larger than the binding energy. In this case, the electron has a large initial kinetic energy directly after ionization. Let us focus on the two example trajectories marked by $\alpha$ and $\beta$ in Fig.~\ref{figopti}b. Both are ionized at instants separated by a small time difference $\delta t\I{i}$. Because $\alpha$ experiences the laser field a little earlier, a velocity difference arises between both.  Note that the velocity difference between $\alpha$ and $\beta$ is conserved in time for a homogeneous laser field.  With a convenient choice of the parameters, the velocity difference between both acquired during $\delta t\I{i}$ can be such that both recollide simultaneously.
In principle, the single-photon ionization process allows for a variety of different ionization directions. However, only the trajectories starting exactly opposite to the polarization direction of the laser recollide and are thus considered. 

The laser field with a shape approaching the ideal one shown 
in Fig.~\ref{figopti}a can be obtained using a small number of Fourier components. The required parameters which are sufficient to reach the  bandwidth-limited pulses below 10~as and 1~as are indicated in Table~\ref{tab_values}.
\begin{table}[h]
\begin{center}
\begin{tabular}{c|c|c|c|c|c|c}
$N\I{F}$&$I\I{L}$[W/cm$^2$]&$\omega\I{X}$[eV]&$I\I{X}$[W/cm$^2$]&ion&$I\I{p}$ [a.u]& $\Delta t$ [as]\\
\hline
8&$10^{16}$&218&$3.5\times 10^{14}$&Li$^{2+}$&4.5&8\\
\hline
20&$10^{17}$&996&$1.4\times10^{15}$&Be$^{3+}$&8&0.8\\
\hline
\end{tabular}
\caption{$N\I{F}$ represents the number of Fourier components contained in the tailored driving pulse, $I\I{L}$ its peak intensity, $\omega\I{X}$  the x-ray frequency employed for ionization, $I\I{X}$ its intensity, $I\I{p}$ the ionization energy and $\Delta t$ the harmonic pulse duration.}\label{tab_values}
\end{center}
\end{table}
However, the method still has several drawbacks. It is experimentally demanding to create a pure ionic gas and to achieve phase-matching in a macroscopic medium due to the free-electron dispersion.  Moreover, the ionization rate is small due to $\omega\I{X}\gg I\I{p}$, and the required large initial momentum and the dipole angular distribution of the ionization process lead to an increased spread of the ionized wave packet as compared to tunnel ionization. Additionally, precise shaping of intense driver fields  with intense harmonics as well as the synchronization of the x-ray and IR pulses is also experimentally demanding.

\section{Experimental applications} \label{EXP-APPLICATION}
Continuously gaining momentum from the increased knowledge on HHG and the co-evolution of technological capabilities, the scientific applications of attosecond HHG light have begun and keep radiating out into a range of various research directions.  Owing to their short-pulsed nature and high photon energies available for probing, the old scientific dream of tracking and watching the dynamics of individual and multiple electrons on their quantum paths throughout the valence and core-shells of atoms and molecules was rekindled.  Using HHG-based sources, the scientific community has already begun to observe the first valence-shell wavefunction dynamics in atoms~\citep{GOULIELMAKIS:RO-10} and molecules~\citep{SMIRNOVA:HI-09}, and will soon explore wavepackets of two~\citep{Argenti:IB-10} or multiple electrons such as in plasmon excitations~\citep{PFEIFER:TR-08,SUESSMANN:AN-11} in motion.  While the femtosecond revolution already opened up the now well-established and fruitful field of femtochemistry (Ahmed Zewail, Nobel prize 1999~\citep{ZEWAIL:NP-00}) making intramolecular motion of atoms accessible to experimentalists, attosecond pulses now promise to be key to monitoring individual bonds in molecules being made or broken~\citep{NUGENTGLANDORF:UF-01,WERNET:RT-09,WOERNER:CR-10}, directly clock the instant of photoionization~\citep{CAVALIERI:CM-07,ECKLE:AS-2008,SCHULTZE:DP-10} or to follow electronic excitations throughout small molecules~\citep{REMACLE:AC-06}.  The experimental capabilities in terms of temporal resolution have come far enough to ask truly fundamental physical questions: Are the physical concepts we use for defining "times" (such as ionization or tunneling time) valid?  Are some questions about the "timing" of electronic processes even ill posed and rather addressed by the definition of phases?  How does the Coulomb potential surrounding atomic and molecular ions, modify the interpretation of measured timings?  And to what extent does the presence of a moderately strong laser field, which typically is involved in measurements, modify the timing of some such fundamental processes?  Here, we will outline a few selected, and by no means exhaustive, frontiers of current experimental approaches invented to measure attosecond electronic quantum dynamics.  What is common to all these methods is the now old-fashioned concept of pump--probe spectroscopy: a temporally compact trigger event (pump), which can be an attosecond pulse or an electric-field extremum within an intense optical laser cycle and a similarly generated probe event, which can either be controlled in time or the temporal position of which can be measured.

\emph{a) Photoelectron spectroscopy methods for time-resolving ionization dynamics} \newline
The first and still most widespread experimental method for time-resolving attosecond processes is the detection of photoelectrons.  Owing to its typical photon energy in the vacuum ultraviolet (VUV) up to the soft-x-ray region, HHG-light interaction with neutral matter almost exclusively results in ionization of the system under study.  The first experiments (see Section~\ref{EXP-SOURCE}) on the characterization of attosecond pulses proceeded through the observation of energy-resolved photoelectrons as a function of delay time between the attosecond and a coherently locked femtosecond pulses, both in the form of attosecond pulse trains (RABBITT~\citep{PAUL:OT-01}) or isolated attosecond pulses (streaking~\citep{HENTSCHEL:AM-01,Itatani:AS-02}).  While for the RABBITT technique, the so-called atomic phase arising from single-photon ionization (a system- and initial-state specific photon-energy $\omega$ dependent phase $\varphi(\omega)$ of the bound-free dipole matrix element) was always included to interpret the results, the streaking technique initially assumed an "instantaneous" response ($\Delta \tau= \frac{\D\varphi(\omega)}{\D \omega}\approx0$) of the photoionization process, i.e. the working principle that the photoionization rate follows the exact intensity shape of the attosecond pulse.  Also due to this assumption, experimental results on the ionization of solids~\citep{CAVALIERI:CM-07} came as a surprise, where it was noticed, by streak-field cross-correlation of a 91~eV attosecond photoionizing pulse with a moderately intense near-visible pulse that electrons seem to ionize "sooner" or "later" depending on their initial state.  The photoionization time difference between the two initial states being either in the valence band or a more localized 4d inner-valence state was measured to be 110$\pm$70~as, and tentatively interpreted as being caused by the different band structure and different transport processes of the electrons on their way to the surface, where they are further accelerated or decelerated by the near-visible laser field for streak-field detection.  Other experiments followed on the photoionization of gas-phase atoms, where similar temporal shifts of photoionization times were discovered.  For Ne ionization by an attosecond pulse, a 20~as "delayed emission" of the 2p vs. the 2s states was found by the same experimental method and tentatively discussed in the context of electron correlation~\citep{SCHULTZE:DP-10}.  As the strong-field approximation was used in reconstructing the temporal shifts, and also as the moderately intense laser pulse modifies the dynamics~\citep{KLUNDER:PS-11,IVANOV:SC-11}, these and the previous solid-state results are still subject of active and controversial discussions~\citep{LEMELL:SS-09,ZHANG:AP-09,BAGGESEN:PE-10,KHEIFETS:DA-10,NAGELE:TR-11}.  It is certain, however, that this experimental approach has reached a level of sophistication that tiny temporal or phase shifts can be extracted, where it is now up to advanced theory to interpret the results of such measurements by inclusion of the Coulomb potential, the probing laser field and the already well-known Eisenbud-Wigner-Smith time delay~\citep{IVANOV:SC-11,WIGNER:LL-55,SMITH:LT-60}, to extract additional knowledge on atomic or molecular~\citep{HAESSLER:PR-09,CAILLAT:AR-11} systems beyond the single-active electron picture.


\emph{b) HHG recollision spectroscopy for measuring electronic wavepackets in molecules} \newline
Instead of measuring photoelectrons or photoions produced in the interaction of HHG attosecond pulses with matter, the HHG process itself can yield dynamical information about electron motion within quantum systems.  With traditional HHG being due to interference of the laser-accelerated recolliding continuum states with the bound initial state of the electron (see Section~\ref{sec:intf-model}), the spectrum of the HHG light contains information, even dynamical information, on the shape of the initial molecular orbital ionized by the strong laser field.  By analyzing HHG from impulsively (field-free) aligned molecules and by comparison of the alignment-angle-dependent HHG spectra to a suitable atomic reference target, the highest-occupied-molecular-orbital (HOMO) shape of nitrogen was reconstructed in a tomographic way~\citep{Itatani:TO-04}.  Already in this first work, the potential of this method with respect to observing electronic wavepacket dynamics was pointed out.  After first signs of the contribution of several molecular orbitals (e.g. HOMO and HOMO-1) to the HHG process~\citep{MCFARLAND2008}, coherent interference between these channels was measured in amplitude and phase and analyzed in dependence of alignment angle, HHG photon energy and driving laser intensity.  From this analysis, the wavefunction shape of the HOMO/HOMO-2 superposition at the time of tunnel-ionization in the HHG process was identified~\citep{SMIRNOVA:HI-09}.  In these experiments, one makes use of the fact that time-resolved information is encoded in the harmonic spectrum, as a certain recolliding (probe) electron energy (and thus emitted photon energy) corresponds to a certain time-delay after the tunnel-ionization (pump) by an electric field maximum.  Additional measurement of the ellipticity of the HHG light as a function of alignment angle and photon energy recently revealed information on the dynamical hole motion after strong-field ionization of nitrogen molecules, including sub-cycle inter-orbital population transfer within the exciting laser pulse~\citep{MAIRESSE:HS-10}.  Another recent implementation of this spectroscopy method allowed insights into multi-electron processes in xenon atoms~\citep{SHINER:PC-11}.


\emph{c) Attosecond transient absorption for observing bound-electron wavepackets} \newline
While the photoelectron/-ion and HHG spectroscopy methods discussed above crucially require continuum electrons, either directly as observables or as an intermediate state to act as a probe interfering with the bound states to be analyzed, a third spectroscopy approach---transient-absorption spectroscopy (TAS)---is not dependent on free electrons and particularly applicable for the spectroscopy of bound--bound transitions.  Instead of measuring electrons, it measures the spectrally resolved absorption (e.g. absorption lines but also non-resonant absorption) of an attosecond pulse transmitted through a coherently excited sample (e.g. by an intense laser pulse), at different controlled time delays.  TAS thus does not rely on intermediate or final continuum electronic states for dynamical probing, often causing problems in interpretation as these states are susceptible to both intense laser fields (typically part of the pump or probe process) as well as the long-range Coulomb field of the ions that are produced.  Continuum electron motion typically does not separate into a Coulomb-only and laser-only perturbation and thus limits direct access to bound--bound dynamics and matrix elements without heavily relying on theory and the validity of suitable models.  Photoelectron detection techniques also suffer from electron backgrounds produced by the strong laser fields alone or by secondary electrons, which do not carry information about the dynamics probed by the combined nonlinear action of both the temporally separated strong laser field and the weak attosecond pulse.  TAS is already an established technique in the femtosecond domain, but was only recently transferred to gas-phase applications with attosecond HHG light sources~\citep{LOH:QS-07}.  Using this method, a long-standing scientific question on the coherence of different spin-orbit hole states populated in strong-field ionization of atoms~\citep{ROTTKE:SP-96} could recently be answered~\citep{GOULIELMAKIS:RO-10}.  In this study, a gaseous but dense (80~mbar) Kr atom sample was strong-field ionized by a an intense (3$\times 10^{14}~$W/cm$^2$) sub-4-fs optical laser pulse, promoting the atoms to the 4p$_{1/2}^{-1}$ and 4p$_{3/2}^{-1}$ ionic states, with one electron missing from either one of the two spin-orbit split states of the 4p shell.  A time-delay controlled attosecond pulse exhibiting photon energies around 80~eV was then sent through the Kr sample to probe these states by transitions of electrons out of the two spin-orbit-split 3d shell states into the two hole states of the 4p shell, resulting in three transitions which are energetically located near 80~eV.  Characteristic spectral absorption lines were thus observed in dependence of the attosecond pulse time delay.  Two out of the three transitions end up in the same final state, making it possible to probe the coherence of the strong-field populated 4p-hole states.  A dynamical oscillation of absorption strength was observed with a period of 6.3$\pm 0.1$~fs, corresponding to the inverse of the energy splitting of these two states, proving their coherent excitation during the strong-field pump process.  Furthermore, it was possible to quantitatively extract the degree of coherence of the excitation, resulting in a value of $0.63\pm 0.17$, relatively close to its maximum of unity as a result of the short ionization pulse that was used.  Recent implementation of light-field synthesis methods, by which the ionization pulse could be further compressed~\citep{WIRTH:ST-11}, allowed the temporal reconstruction of the spin-orbit wavepacket in real time relative to the inducing laser field.  Other successful implementations of transient-absorption spectroscopy for the measurement of electron dynamics include the laser-induced soft-x-ray transparency by coupling doubly-excited electronic states~\citep{LOH:FI-08} and measurement of electron wavepacket interference~\citep{HOLLER:WI-11} in singly-excited He atoms, and the time-domain observation of autoionizing states after inner-shell excitation in Ar atoms~\citep{WANG:Ar-10}.

\section{Outlook}
After more than two decades of HHG, a profound understanding of the underlying mechanism has been reached and numerous applications have been put forward. With respect to the latter, attosecond science has been dominating and the notion of visualizing atomic structures in motion has been materialized in most beautiful manner. Still numerous challenges remain: The transfer of HHG to the multi-keV and MeV range and consequently the generation of zeptosecond pulses represents a challenge. In spite of significant progress, laser-induced relativistic atomic dynamics still needs to be optimized with regard to larger recollision and recombination yields and consequently stronger HHG signals. In addition to visualizing nuclear and high-energy
processes another enormous challenge is the time-resolved spectroscopy of more complex systems like e.g. biomolecules. With alternative facilities like free-electron lasers and plasma-based schemes becoming increasingly relevant, atomic HHG light sources are likely to remain at least competitive due to their unmatched degree of spatial and temporal coherence, including the synchronization of the generated attosecond pulses to optical fields, and due to to their availability in numerous laboratories worldwide at reasonable expense.   

%

\bibliography{ref_review_hhg_EA}

\end{document}